\documentclass[12pt]{article}
\usepackage{amssymb}
\usepackage{graphicx,psfrag}
\usepackage{multicol}
\usepackage{journals}

\newcommand{\bea}{\begin{eqnarray}}
\newcommand{\eea}{\end{eqnarray}}
\newcommand{\simgt}{\hbox{ \raise3pt\hbox to 0pt{$>$}\raise-3pt\hbox{$\sim$} }}
\newcommand{\simlt}{\hbox{ \raise3pt\hbox to 0pt{$<$}\raise-3pt\hbox{$\sim$} }}

\begin{document}

\def \etal     {\relax\ifmmode{et \; al.}\else{$et \; al.$}\fi}
\newcommand{ \sla}    {\hspace*{-1.8mm}/\hspace*{0.1mm}}

\thispagestyle{empty}
\vspace*{-3.5cm}
\begin{flushright}
\today\\
TU--759
\end{flushright}

\vspace{0.5in}

\begin{center}
  \begin{LARGE}
  {\bf Search for Anomalous Couplings in Top Decay at Hadron Colliders}
  \end{LARGE}
\end{center}

\vspace{0.4in}

\begin{center}
S. Tsuno$^{a}$,
I. Nakano$^{a}$, Y. Sumino$^{b}$, and R. Tanaka$^{a}$
\end{center}

\begin{center}
\emph{$^{a}$Department of Physics, Okayama University, 
        Okayama, 700-8530, Japan} \\ \vspace{1.0ex}
\emph{$^{b}$Department of Physics, Tohoku University,
        Sendai, 980-8578, Japan} \\ \vspace{1.0ex}
\end{center}

\vspace{1.truein}

\begin{abstract}
We present a quantitative study on sensitivities to 
the top-decay anomalous couplings,
taking into account realistic experimental conditions expected at Tevatron
and LHC.
A double angular distribution of $W$ and charged lepton in the top decay
is analyzed,
using $t\bar{t}$ events in the $lepton+jets$ channel.
In order to improve sensitivities to the anomalous couplings,
we apply two techniques:
(1) We use a likelihood fitting method for full kinematical reconstruction of
each top event.
(2) We develop a new effective spin reconstruction method for
leptonically-decayed top quarks; this method
does not require spin information of the antitop side.
For simplicity, we neglect couplings of right-handed bottom
quark as well as $CP$ violating couplings.
The 95\% C.L. estimated bound 
on a ratio of anomalous couplings reads
$-0.81 < f_2/f_1 < -0.70$, $-0.12<f_2/f_1<0.14$ using 1000 reconstructed
top events at Tevatron, while
$-0.74<f_2/f_1<-0.72$, $-0.01<f_2/f_1<0.01$ is expected with
100k reconstructed top events at LHC,
where only statistical errors are taken into account.
A two-fold ambiguity in the allowed range remains when the number of events
exceeds a few hundred.

\end{abstract}

\vspace{0.1in}
\newpage

\section{Introduction}

The top quark is the heaviest elementary particle discovered up to now.
Namely, the top quark mass term breaks the electroweak gauge symmetry
maximally among all of the observed interactions of elementary particles.
For this reason, we 
expect that the top quark can be used as a probe to search into the
symmetry-breaking physics.
On the other hand, so far reported experimental data from Tevatron 
on top quark properties are still limited;
see e.g.\ \cite{topmass,RatioBR,Whelicity}.\footnote{
In addition, there are some indirect constraints on anomalous $ttZ$ and $tbW$
interactions from the precision measurements at LEP 
and from the 
flavor-changing-neutral-current (FCNC) decays of the bottom quark 
\cite{Hosch,Hikasa,const2}.
}
No sign of significant deviations from the Standard Model (SM) predictions
has been seen.

The number of observed top quark events in the Tevatron Run II experiment
is increasing steadily and now reaching of the order of a few hundred.
Moreover, it is expected that at LHC experiment an immense number of
top quarks
will be produced.
Thus, we foresee that detailed properties of the top quark will
start to be uncovered in near future.


Among various interactions of the top quark, study 
of the top quark decay properties is particularly interesting.
In the SM (and many of its extensions), the
top quark decays via electroweak interaction before hadronization.
Hence, the top quark's spin information is transferred directly to its
decay daughters, and their distributions can be predicted reliably
using perturbative calculations.
Thus, the top quark spin can be used as a powerful analysis tool
for scrutinizing top quark interactions.

There have been many theoretical studies on how to test
top quark decay properties at hadron colliders.
Non-standard effects on the total
top decay width and on the top decay branching fractions to polarized $W$ states
have been computed in the minimal-supersymmetric standard model \cite{WidthSUSY,Cao},
in a $R$-parity violating supersymmetric model \cite{Nie}, and in the top-color
assisted technicolor model \cite{Wang}.
A study was given on
how to extract anomalous $tbW$ couplings and
discriminate different underlying models from 
combined measurements of 
the top decay branching fractions to polarized $W$s and the 
single top production rates \cite{Chen}.
The top quark decay $t \to bW$ was studied within 
the non-commutative standard model in \cite{noncomm}.
The correlation between $t$ and $\bar{t}$ spins in the
$t\bar{t}$ events has been studied as a mean to
investigate decay properties of the top quark 
\cite{corr1,angularcorr,Stelzer,offdiagonal,corr4}.
Effects of anomalous couplings on the lepton
rapidity and transverse energy distribution were briefly discussed in \cite{Lampe}.
There have also been a vast number of
studies on top rare decays and non-standard
decay channels; see \cite{Beneke} and references therein.

In this paper we focus on effects of the anomalous $tbW$ couplings to
the distributions of $W$ and charged lepton in the decay of top quarks.
We estimate sensitivities to (some of) these couplings expected at Tevatron and LHC,
using Monte Carlo simulations
which take into account realistic experimental conditions at
these colliders.
There have been no other
quantitative studies, using top quark decays, 
on sensitivities to the top decay anomalous couplings
at hadron colliders.

A sensitivity study of the anomalous couplings using
the single top production process was given in \cite{SensitivitySingleTop}.
Their estimated sensitivities to the anomalous couplings are rather low, 
due to existence of huge background cross sections for $Wb\bar{b}$ and
$Wb\bar{b}+jets$ processes.
The small signal-to-noise ratio leads to large statistical 
as well as systematic errors.
In this connection, we note that due to lack of data statistics and difficulty 
in the background estimation, the single top production 
process has not yet been observed at Tevatron \cite{singletopcdf,singletopd0}.
As compared to their analysis, our analysis method using $t\bar{t}$
events in the $lepton+jets$ mode
is cleaner and
involves controlled and small backgrounds.
Consequently, our method improves sensitivities to the anomalous couplings
considerably as compared to the results of \cite{SensitivitySingleTop}.

In our analysis, we assume that there are only anomalous $tbW$ 
couplings for the left-handed
bottom quark and we neglect $CP$ violation.
We impose these assumptions, for simplicity of our analysis, 
and also in view of the present bounds on the general $tbW$ couplings.
This leaves only two independent real anomalous couplings,
and our analysis is sensitive only to the ratio of these two couplings.
See Sec.~2 for details.

In order to improve sensitivities to the anomalous couplings, 
we devise two techniques.
(1) We use a likelihood fitting method
for full kinematical reconstruction of each $t\bar{t}$ event.
(2) We develop a new method for reconstructing an effective spin direction
of the top quark.
In particular, the new feature of the latter technique is that we do not
need to reconstruct the spin of the antitop side in the $t\bar{t}$ events,
i.e.\ we do not make use of the correlation between the top and antitop spins.
In a separate paper, two of the present authors elucidate
theoretical aspects of
the effective spin reconstructed in this method
\cite{letter}.

As well known, top quarks produced at Tevatron and LHC are
scarcely polarized.
It is one of the major
reasons why people have considered correlations between
top spin and antitop spin in the $t\bar{t}$ production process
\cite{corr1,angularcorr,Stelzer,offdiagonal,corr4};
with the help of this correlation, one can, in principle, reconstruct the information 
of the top quark spin, by looking into the information on the antitop side.
Nevertheless, if we are to use events in
the {\it dilepton} channel, the event statistics is rather
low (especially at Tevatron), leading to disadvantages with regard to the
sensitivity study.
On the other hand, if we are to use events in the $lepton+jets$ channel (which has not been
tried at Tevatron up to now), we suffer from large systematic uncertainties
due to the complexity in the reconstruction of 
the spin of a hadronically-decayed
top quark.

At first our claim may seem unreasonable, as
one might argue that it is impossible to reconstruct the spin of an 
unpolarized top quark:
Since an unpolarized state is rotationally invariant, there exists no reference
direction appropriate for a spin direction.
While this argument is correct on its own,
we can still reconstruct an effective ``spin direction'' of a top quark,
practically useful in the analysis of top decay, in the following sense.
Unpolarized top quarks can be interpreted as an admixture, where one-half of
them have their spins in $+\vec{n}$ direction and the other half have
their spins in $-\vec{n}$ direction, for an arbitrary
chosen unit vector $\vec{n}$.
Then the directions of the decay products from the top quarks
with $+\vec{n}$ spin are strongly correlated with the $+\vec{n}$
direction, provided the top decay interaction is close to the SM prediction.
For instance, the charged leptons are emitted preferentially in 
$+\vec{n}$ direction in the rest frame of the top quark.
The same is true for the top quarks with $-\vec{n}$ spin.
Then it seems reasonable (at least intuitively) to project the 
direction of the lepton $\vec{n}_l$ onto the $\vec{n}$-axis
and define an effective spin direction as 
${\rm sign} (\vec{n}\!\cdot\!\vec{n}_l)\times\vec{n}$,
for each event.
Indeed certain angular distributions of the top 
decay products with respect to this effective spin direction
reproduce fairly well the corresponding angular distributions from a
truly polarized top quark.
This is the case even including anomalous couplings.
It is in this sense that the effective spin direction is practically useful.
That we can choose any axis $\vec{n}$, and that any choice
is equivalent (if we ignore experimental environment), guarantee the rotational
invariance of the unpolarized state of the top quark.

In Sec.~2 we present our theoretical setups, namely the definitions of the
anomalous couplings and theoretical formulas for the decay angular distributions.
In Sec.~3 we propose our method for reconstructing an effective top
spin direction and discuss why it can be useful.
The MC simulations used in our analysis are explained in Sec.~4.
Sec.~5 demonstrates the kinematical reconstruction of $t\bar{t}$ events
using our likelihood fitting method.
Sensitivities to the anomalous couplings are estimated using the
selected $t\bar{t}$ event samples in Sec.~6.
Conclusions are given in Sec.~7.
In Appendix we present the theoretical formula for the double angular
distribution when we use the effective spin direction.

\section{Anomalous couplings in top decay vertex}
\label{s2}

It is conventional to incorporate
effects of physics beyond the SM in various form factors
of the interactions among the known particles. 
The interactions of
fermions and gauge boson, in general, can be expressed by six form factors with a 
particular energy scale at which new physics is opened. 
If we assume that $W$ boson is on-shell, the number of 
the form factors is reduced to four. 
Thus, the effective $Wtb$ vertex relevant to the top quark decay is expressed 
as \cite{cpyuan}
\begin{equation}
\Gamma_{Wtb}^{\mu} = - \frac{g}{\sqrt{2}} V_{tb} 
\bar{u}(p_{b})
\left[
\gamma^{\mu}(f_{1}^{L}P_{L} + f_{1}^{R}P_{R})
-\frac{i \sigma^{\mu\nu}k_{\nu}}{M_{W}}(f_{2}^{L}P_{L} + f_{2}^{R}P_{R})
\right]
u(p_{t}) \quad ,
\label{eq1}
\end{equation}
where $V_{tb}$ is the 
CKM (Cabibbo-Kobayashi-Maskawa)  matrix element \cite{ckm}, 
$P_{L,R}$ $=$ 
$(1 \mp \gamma_{5})/2$ is the left-handed/right-handed projection operator, 
and $k$ is the momentum of $W$. 
We take the convention in which the energy scale is represented by $M_{W}$. 
The form factors $f_{1}^{L,R}$ and 
$f_{2}^{L,R}$ are in general complex. 
At tree-level of the SM, their values are 
$f_{1}^{L}=1$ and $f_{1}^{R}=f_{2}^{L}=f_{2}^{R}=0$.
We will be concerned only with  the top quark decay process $t \to b W$, where 
$Q^2$ value is fixed, therefore, we treat
the form factors as constants (couplings) henceforth.
The decay vertex for the antitop quark can be written similarly
in a straightforward manner.

There are some (indirect) constraints to these couplings from measurements of
the FCNC processes of rare $B$ decays, $b \rightarrow s \gamma$ and 
$b \rightarrow sl^{+}l^{-}$, and from the precision 
measurements at $Z$ pole \cite{const2}. 
These measurements 
support consistency with the SM predictions. 
Apart from a somewhat loose constraint for the $f_{2}^L$ parameter,
the non-Standard 
$CP$-violating and right-handed bottom quark couplings are severely
constrained. 
Taking this into account, and for simplicity of the analysis, 
we assume in the following that the 
interactions in Eq.~(\ref{eq1}) preserve $CP$ symmetry and also neglect the 
couplings of the right-handed bottom quark.
Hence, Eq.~(\ref{eq1}) is reduced to
\begin{eqnarray}
&&
\Gamma_{Wtb}^{\mu} = - \frac{g}{\sqrt{2}} V_{tb} 
\bar{u}(p_{b})\left[ \gamma^{\mu}f_{1}P_{L}
-\frac{i \sigma^{\mu\nu}k_{\nu}}{M_{W}}f_{2}P_{R}\right]
u(p_{t}) \quad ,
\label{eq2}
\\ &&
\bar{\Gamma}_{W\bar{t}\bar{b}}^{\mu} = - \frac{g}{\sqrt{2}} V_{tb} 
\bar{v}(p_{\bar{t}})\left[ \gamma^{\mu}{f}_{1}P_{L}
-\frac{i \sigma^{\mu\nu}k_{\nu}}{M_{W}}{f}_{2}P_{L}\right]
v(p_{\bar{b}}) \quad ,
\label{eq3}
\end{eqnarray}
where 
\begin{equation}
f_{1} \equiv f_{1}^{L} = \bar{f}_{1}^{L} \quad , \quad
f_{2} \equiv f_{2}^{R} = \bar{f}_{2}^{L} \quad ,
\label{eq4}
\end{equation}
 and both $f_1$ and $f_2$ are real. 
For definiteness, we have shown both the top and antitop decay vertices 
explicitly.
As stated, we have simply neglected $f_{1}^{R}$($\bar{f}_{1}^{R}$) and 
$f_{2}^{L}$($\bar{f}_{2}^{R}$) terms.
Nevertheless,
even if they happen to be non-vanishing (but not large),
their effects are expected to be suppressed, 
since they enter the cross section formulas quadratically in the limit
$m_b \to 0$,
as they do not interfere with the SM amplitude.
Hence, our treatment would be justified for a first analysis.

\begin{figure}[t]
\begin{center}
\includegraphics[width=8.0cm]{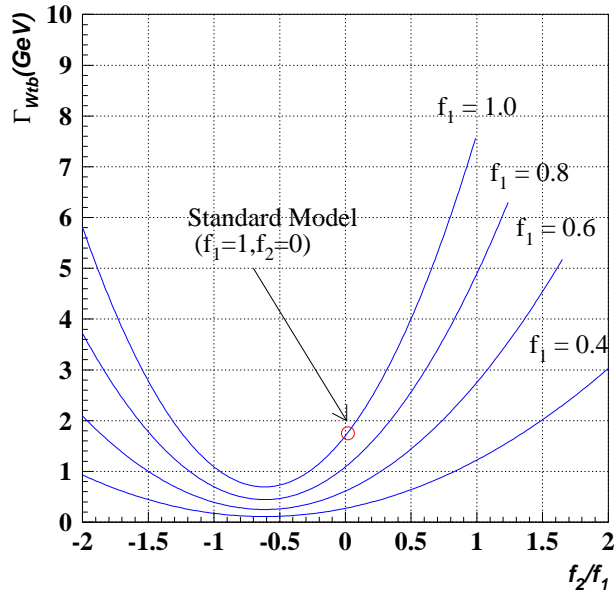}
\caption{\small 
Tree-level partial width for $t \to b W$ as a 
function of $f_2/f_1$ and for different values of $f_1$.
The unit is in GeV. 
The tree-level SM prediction corresponds to 
($f_{1}$,$f_{2}$)=(1,0).}
\label{tpprtlwid}
\end{center}
\end{figure}

In Fig.~\ref{tpprtlwid}, we show the partial decay width for $t \to b W$
for different values of $f_1$ and  $f_{2}$/$f_{1}$, where the 
tree-level SM corresponds to 
($f_{1}$,$f_{2}$) = (1,0). 
Apart from the overall normalization proportional to $f_1^{\, 2}$,
the partial decay width is a quadratic function of $f_{2}$/$f_{1}$.
One sees that the partial decay  width is below
10~GeV in a wide region in the ($f_{1}$,$f_{2}$) parameter space.
The current resolution of the reconstructed top-quark invariant mass 
distribution using jet events at Tevatron is order 
40~GeV.
It follows that a wide region in the parameter space ($f_{1}$,$f_{2}$) is
still allowed under the constraint from the present top invariant mass
measurement.

In principle, we can use the present measurement of $W$ helicities 
\cite{Whelicity} for
constraining $f_2/f_1$.
No explicit analysis has been given so far, however.
From a rough estimate, we conjecture that
a range $|f_2/f_1| \simlt 0.3$ is (at least) scarcely
constrained at the present status.

Let us discuss effects of the anomalous couplings, according to
Eqs.~(\ref{eq2}) and (\ref{eq3}), on
the distribution of the decay products from the top quark.
We may separate the dependences of the decay distribution
on $f_1$ and on $f_2/f_1$.
A variation of $f_1$ changes only the normalization of the
(partial) decay width of the 
top quark, while a variation of $f_2/f_1$ changes both the normalization and the 
differential decay distributions.
Since it is difficult to measure the absolute value of the decay width 
accurately in near future, 
our primary goal will be to measure (constrain) the value of $f_2/f_1$
from the measurement of the differential decay distribution.
An efficient method was proposed in \cite{ikematsu} using 
the decay process of the top quark
at future $e^+e^-$ collider experiments.
Noting that only the decay process is concerned, we can apply the main strategy of
their method to our analysis aimed for hadron collider experiments.
The relevant strategy is as follows.
Since the transverse $W$ boson (denoted as $W_{T}$) is more 
sensitive to $f_{2}$ than the longitudinal $W$ boson ($W_{L}$), we can enhance
the contribution of $W_{T}$ using the decay distribution.
It is well known that the contribution of $W_{T}$ is dominant when
$W$ is emitted opposite to the top spin direction
in the decay $t \to bW$ and also when $l$ is emitted in the opposite
direction to $W$ in the decay $W \to l\nu$.
Hence, we can select 
these kinematical regions in order to enhance sensitivity to
$f_2/f_1$.

\begin{figure}[t]
\begin{center}
\psfrag{AW}{$W$}
\psfrag{Atop}{top}
\psfrag{Aspin}{spin}
\psfrag{WRestFrame}{$W$ rest frame}
\psfrag{ThetaW}{$\theta_{W}$}
\psfrag{ThetaL}{$\theta_{l}$}
\psfrag{PhiL}{$\phi_{l}$}
\psfrag{eL}{$l$}
\includegraphics[width=5.0cm]{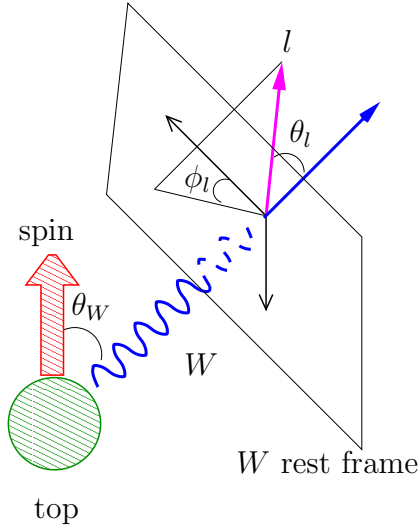} 
\caption{\small Schematic view of the angles used in Eq.~(\ref{eq5}). 
}
\label{angle_def}
\end{center}
\end{figure}

The differential decay 
distribution of $W$ and $l$ in the semi-leptonic decay from a top quark 
with definite spin orientation is expressed as \cite{cpyuan}
\begin{eqnarray}
\frac{d\Gamma(t \rightarrow bW \! \rightarrow  bl\nu)}
{d\cos\theta_{W}d\cos\theta_{l}d\phi_{l}} =
A 
{\left|
(f_{2}+f_{1}\frac{M_{t}}{M_{W}})\cos\frac{\theta_{W}}{2}\sin\theta_{l}
+(f_{1}+f_{2}\frac{M_{t}}{M_{W}})e^{-i\phi_{l}}\sin \frac{\theta_{W}}{2}
(1-\cos\theta_{l}) 
\right|}^{2} ,
\nonumber\\
\label{eq5}
\end{eqnarray}
with
\begin{equation}
A = \frac{3G_{F}|V_{tb}|^{2}M_{W}^{2}(M_{t}^{2}-M_{W}^{2})^{2}}
{32\sqrt{2}\pi M_{t}^{3}} \times Br(W \rightarrow l\nu) \quad .
\label{eq6}
\end{equation}
Here, $G_{F}$ is the Fermi constant.
$\theta_{W}$ is defined as the angle between the top spin 
direction and the direction of $W$ in the top quark rest frame. 
$\theta_{l}$ is defined as the lepton 
helicity angle, which is the angle of the charged lepton in the rest frame of 
$W$ with respect to the original direction of the travel of $W$. 
$\phi_{l}$ is 
defined as the azimuthal angle of $l$ around the original direction of 
the travel of $W$.
A schematic view of the angle definitions is shown in 
Fig.~\ref{angle_def}. 
The first term in the amplitude on the right-hand-side of Eq.~(\ref{eq5})
represents the contribution of 
$W_L$, while the second term represents the contribution
of $W_T$.\footnote{
We note that the $\cos\theta_l$ distribution is used 
in the $W$ helicity measurement \cite{Whelicity}.
}

After integrating over $\phi_l$, we obtain the double angular distribution
\bea
&&
\frac{
d\Gamma(t \rightarrow bW \! \rightarrow  bl\nu)
}{
d\cos\theta_{W}d\cos\theta_{l}
} =
\pi \, A \,
\biggl[
\Bigl(f_1\frac{M_t}{M_W}+f_2 \Bigr)^2
\cos^2 \frac{\theta_W}{2} \, \sin^2 \theta_l
\nonumber \\
&& ~~~~~~~~~~~~~~~~~~~~~~~~~~~~~~~~~~~~~~~~~
+ 4 \, \Bigl(f_1 +f_2\frac{M_t}{M_W} \Bigr)^2
\, \sin^2 \frac{\theta_W}{2} \,\sin^4  \frac{\theta_l}{2} 
\biggr] \, .
\label{doubleangdistr}
\eea
The one-loop QCD correction to the distribution Eq.~(\ref{doubleangdistr})
within the SM is known \cite{QCD1Loop}.
A large part of the correction goes to a variation of the normalization
of the partial decay width, which amounts to about 9\%, whereas
the correction to the distribution shape (after the correction to the
normalization is removed)
is at the level of 1--2\% or less.
For simplicity,
in most of the following discussion we discard the effect of the QCD correction;
see also \cite{letter}.

\begin{figure}[hptb]
\begin{center}
\begin{tabular}{ccc}
\includegraphics[width=6.0cm]{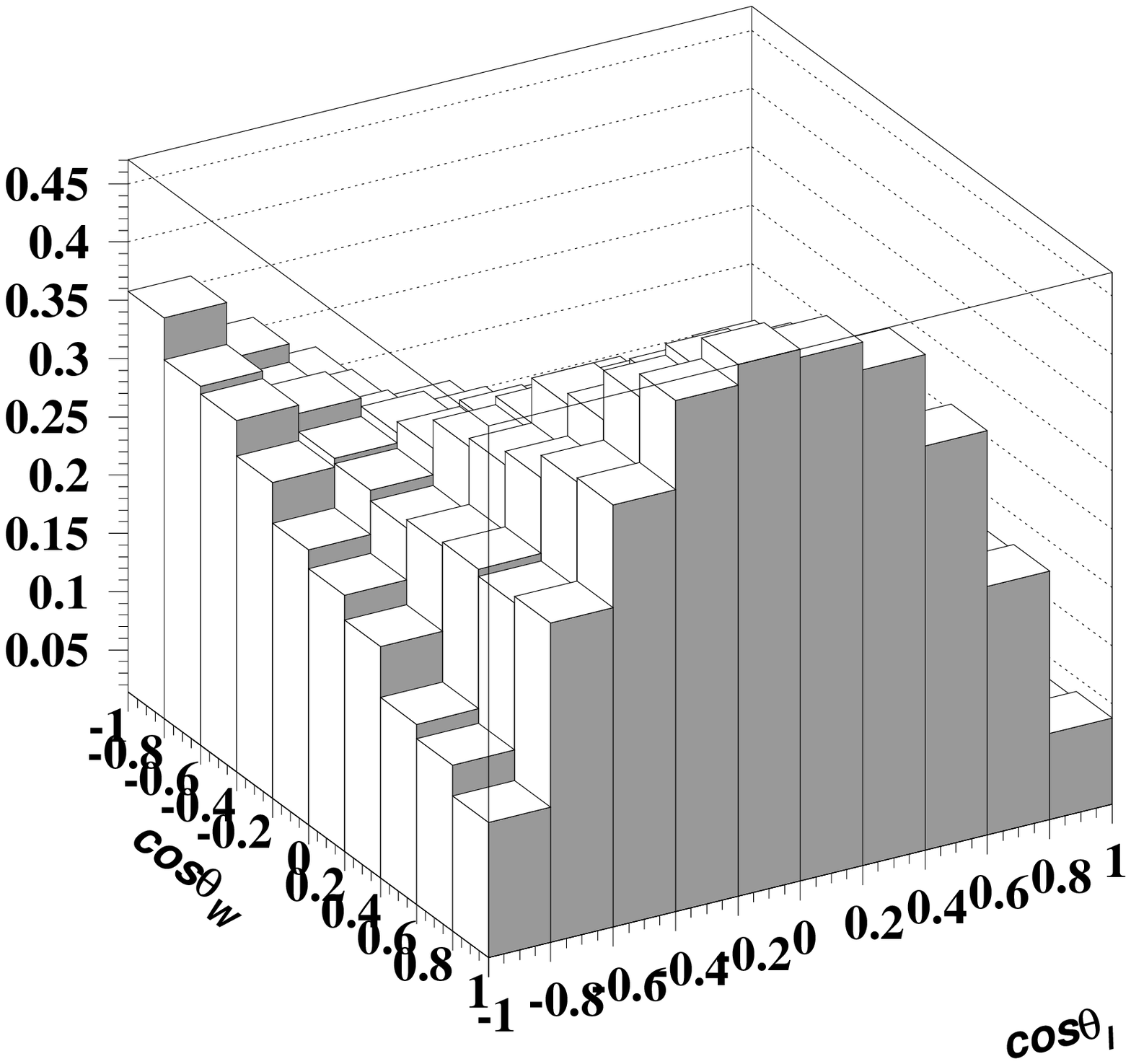} &&
\includegraphics[width=6.0cm]{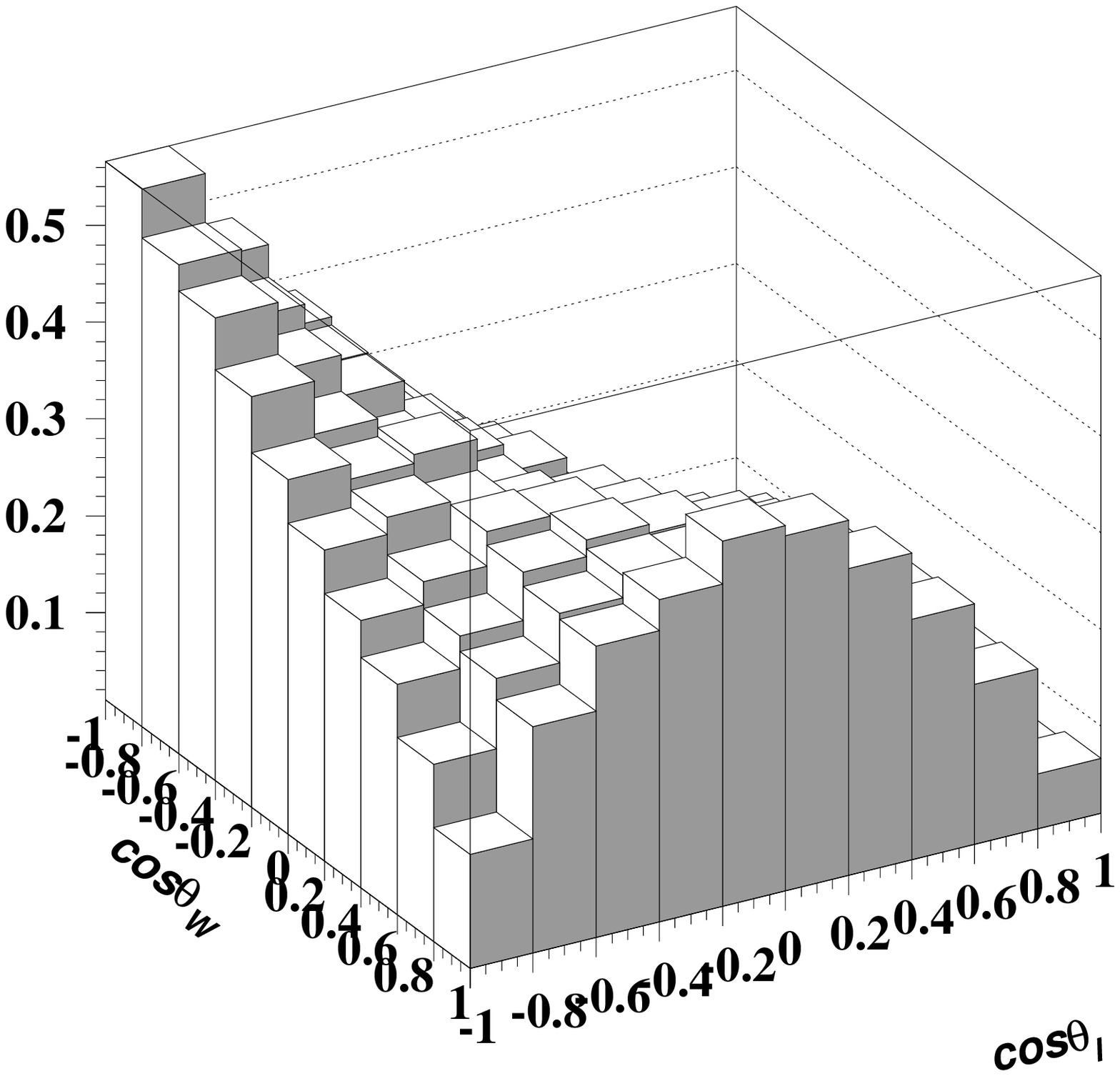}\\
(a) && (b)
\end{tabular}
\caption{\small Normalized differential decay distributions (a) for $(f_1,f_2) =(1,0)$,
and (b) for $(f_1,f_2) =(1,0.3)$.
They are normalized to unity upon integration.}
\label{spin_sm-f103}
\end{center}
\end{figure}

In Figs.~\ref{spin_sm-f103}, we show the normalized double angular distribution 
$N^{-1}d\Gamma(t \rightarrow bl\nu)/d\cos\theta_{W}d\cos\theta_{l}$;
Fig.~\ref{spin_sm-f103}(a) corresponds to $(f_1,f_2) =(1,0)$ (tree-level SM) and
(b) to $(f_1,f_2) =(1,0.3)$, respectively. 
Comparing the two figures,
the effects of varying $f_2$ are indeed enhanced in the regions
$\cos\theta_{W} \simeq -1$ and $\cos\theta_{l} \simeq -1$, in accord
with enhancement of the
$W_{T}$ contributions in these regions.

Thus, it is crucial to reconstruct
 the top quark's spin orientation in this method.
At hadron collider experiments,
it is much more non-trivial to reconstruct the top quark
spin direction, as compared to $e^+e^-$ collider experiments. 
We discuss how to reconstruct the top spin direction in the next section.


\section{Effective spin reconstruction}
\label{s3}

At hadron colliders, top 
quarks are produced dominantly through $t\bar{t}$ production processes.
At Tevatron, 85\% of the produced $t\bar{t}$ pairs come from $q\bar{q}$
initial states, while 15\% come from $gg$ initial states.
On the other hand, at LHC, the corresponding fractions are 10\% ($q\bar{q}$)
and 90\% ($gg$), respectively.
At these colliders, polarization of the produced top quarks is rather small:
at tree level, produced top quarks are unpolarized;
at NLO, polarization of top quarks is reported to be very 
small \cite{nlospin}.
Therefore, a priori, the spin orientation of a produced top quark is
unknown.
This is in contrast to $e^+e^-$ colliders, where sizable polarization of
top quarks is expected due to 
parity-violating nature of the interactions in the top production
process.

In our analysis of the anomalous couplings, we want to utilize correlations
between the top quark
spin direction and the distribution of its decay daughters.
As already mentioned, in most of the existing analyses, people have considered
correlations
between the top spin and antitop spin in the $t\bar{t}$ production
process.
In principle we may use these correlations to reconstruct the spin direction
of the parent top quark.
Namely, we may use the information of the decay distributions on the antitop
side to reconstruct the top spin direction, and then examine the correlation
between the top spin direction and the distribution of its decay products.
A serious deficit of such methods is that they are quite complicated.
For instance, the direction of
the down-type quark in the hadronic decay of antitop quark is maximally
correlated with the antitop spin.
Hence, in order to reconstruct the spin of the antitop quark,
we should distinguish the charges of the quarks from $W$.
This is a highly non-trivial task and we anticipate that rather large systematic
errors will be involved before eventually reconstructing the top quark spin.
On the other hand, if we want to use semi-leptonic decays both on top and antitop 
sides, we suffer from lack of statistics, as well as it is non-trivial to solve
kinematics due to the two missing momenta of the neutrinos.

Here we take another route for reconstructing (effectively) the top quark spin.
We use the correlation between the top spin and the direction of
the charged lepton in the top decay for reconstructing the parent top quark's spin,
and then use it to analyze the decay anomalous couplings of the same top quark.
Since we both reconstruct the spin and analyze the
spin-dependent decay distribution using the {\it  same top decay process},
we should make sure that we use independent correlations in the
former and latter procedures to avoid obtaining
a meaningless outcome.
For this purpose, we take advantage of the following facts:
(1) Within the SM, the charged lepton is 
known to be the
best analyzer of the parent top quark's spin and is 
produced preferentially in the direction of the top spin \cite{jk1}.
(2) The angular distribution of the charged lepton with respect to the
top spin direction (after all other kinematical 
variables are integrated out) is hardly affected by the anomalous couplings of
top quark, if the anomalous couplings are small  \cite{new}.\footnote{
More precisely, $\ell^\pm$ angular distribution is independent of 
the anomalous couplings
up to (and including) linear terms in these couplings.
}
Therefore, we may project the direction of the charged lepton
onto an appropriate spin basis; then
the reconstructed top quark spin
is scarcely affected by existence of the anomalous
couplings $f_1$ and $f_2$ when they are small.
It follows that this spin direction
is useful in testing the differential decay distribution
described in the previous section,
which is sensitive to the anomalous couplings.

Provided that produced top quarks are perfectly unpolarized,
and provided
that we disregard kinematical cuts and acceptance corrections, 
there is no difference on which
spin basis we choose to project the direction of the charged lepton.
Suppose we choose for the basis-axis an arbitrary unit vector $\vec{n}$
in the top rest frame.
Then, if the lepton is emitted on the same side as $\vec{n}$,
i.e.\ if ${\vec{n}\cdot\vec{p}_l}>0$
($\vec{p}_l$ is the lepton momentum in the top rest frame), we define the
``spin vector'' to be $\vec{n}$; on the other hand, if ${\vec{n}\cdot\vec{p}_l}<0$,
we define the
``spin vector'' to be $-\vec{n}$.
The differential decay distribution 
$d\Gamma/d\cos\theta_Wd\cos\theta_l$
with respect to thus defined ``spin vector''
can be computed analytically, which we  present in the Appendix.
\begin{figure}[hptb]
\begin{center}
\begin{tabular}{ccc}
\includegraphics[width=6.0cm]{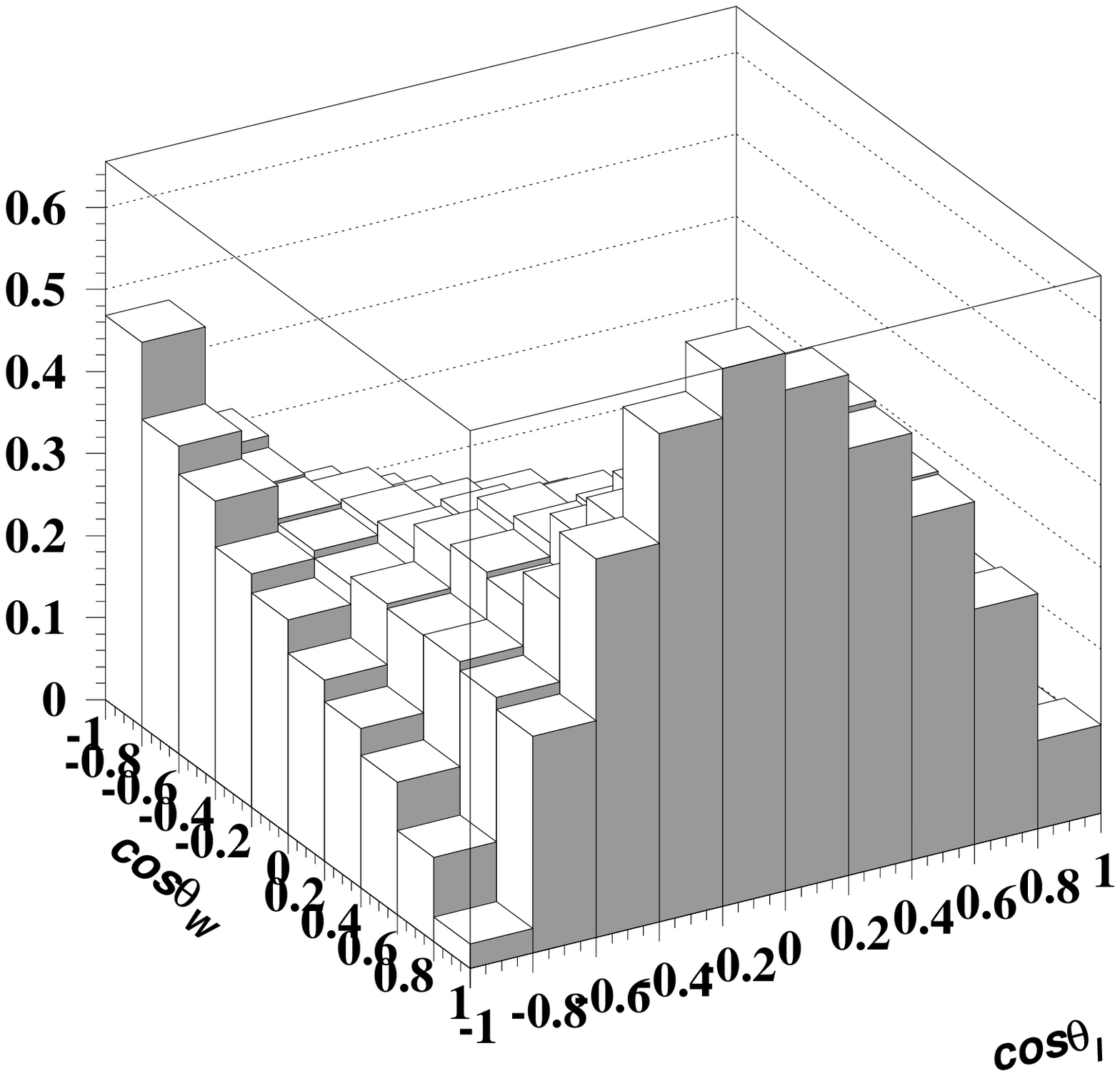} &&
\includegraphics[width=6.0cm]{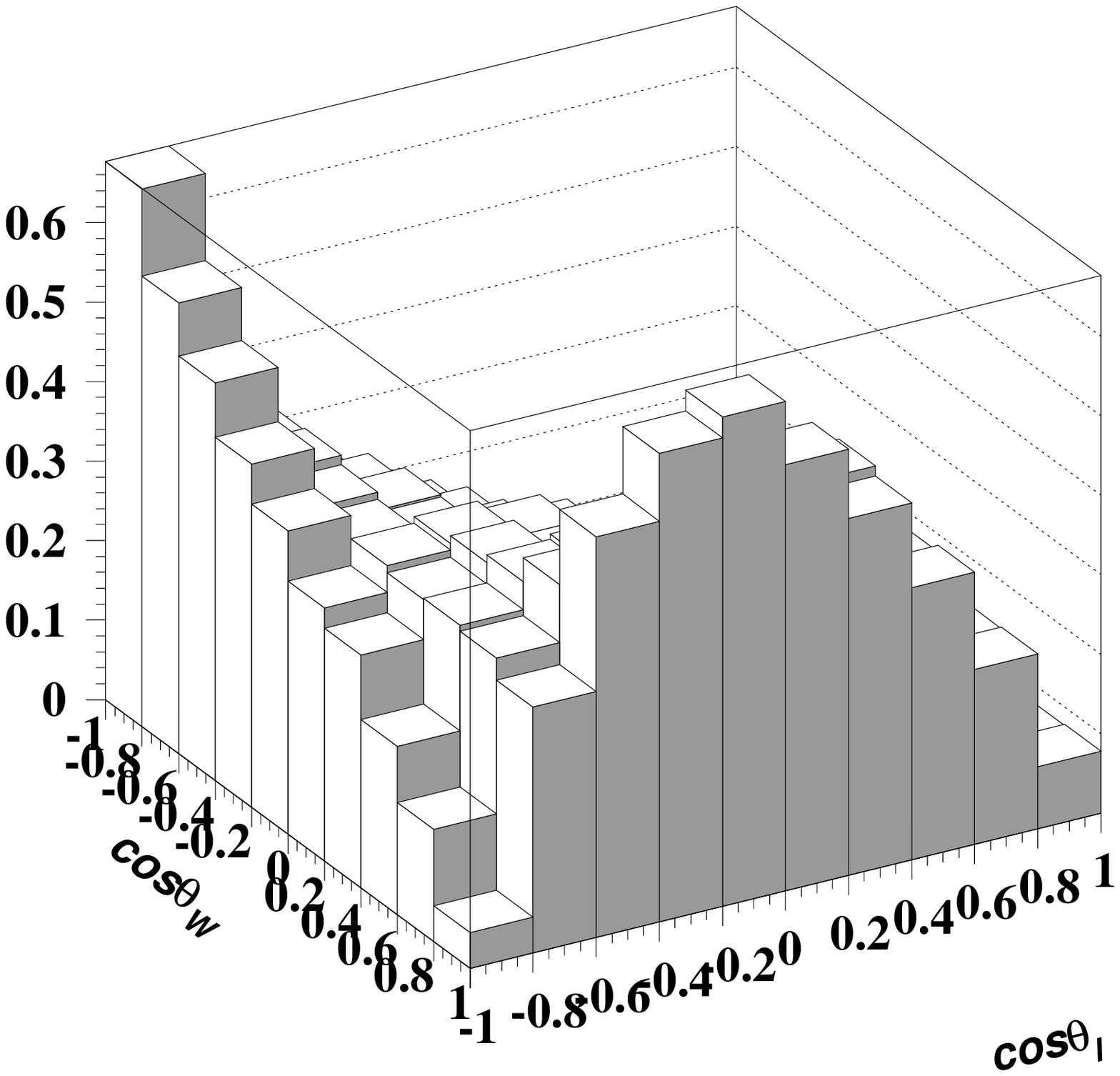}\\
(a) && (b)
\end{tabular}
\caption{\small Normalized differential decay distributions 
using the reconstructed effective spin direction 
${\rm sign}({\vec{n}\cdot\vec{p}_l}) \vec{n}$ for
(a) $(f_1,f_2) =(1,0)$,
and (b) $(f_1,f_2) =(1,0.3)$.
These figures reproduce well the original distributions
in Figs.~\ref{spin_sm-f103}.
}
\label{fin_smgen}
\end{center}
\end{figure}
In Figs.~\ref{fin_smgen},
we show this double angular distribution for $(f_1,f_2)=(1,0)$ and $(1,0.3)$.
One can see that the distributions approximate the corresponding distributions
in Figs.~\ref{spin_sm-f103}, which were computed using the 
spin direction of a polarized top quark.
Qualitative features of the
bulk distribution shape as well as of the dependence on $f_2/f_1$ are
reproduced.
It is this approximation to (reproduction of) the original double angular distribution
that guarantees a good efficiency of our spin reconstruction method
in the analysis of the anomalous couplings.
See \cite{letter} for the study on theoretical aspects of the reconstructed
spin direction in this method.

In practice, if we take into account realistic experimental conditions,
different choice of spin basis (axis) $\vec{n}$ leads to different sensitivities to
the anomalous couplings, due to effects of kinematical cuts.
Here, we examine three types of spin basis that have been 
analyzed in the literature,
then we choose the basis which is most suited for our analysis.

The beamline basis \cite{angularcorr} is to take the spin axis
($\vec{n}$) to be either of the initial beamline
direction in the $t\bar{t}$ c.m.\ frame;
the helicity basis is to take $\vec{n}$ to be the direction of top quark
in the $t\bar{t}$ c.m.\ frame;
the off-diagonal basis is defined to be a linear combination of
the former two bases in such a way to maximize the correlation
between the $t$ and $\bar{t}$ spins \cite{offdiagonal}.\footnote{
The original definition of the beamline basis (off-diagonal basis)
is given in the top rest frame (zero-momentum frame of
the initial partons).
For convenience of our analysis,
we redefine in the $t\bar{t}$ c.m.\ frame.
}
Advantages and disadvantages of these bases have been studied
in the context of spin correlations between $t$ and $\bar{t}$
in the $t\bar{t}$ production processes
\cite{corr1,angularcorr,Stelzer,offdiagonal,corr4}.\footnote{
For $q\bar{q}$ initial state,
the spin correlation is 100\% in the off-diagonal
basis.
The spin correlation in the beamline basis is somewhat
smaller but considerably larger than
that in the helicity basis.
}
Nevertheless, these are irrelevant in our spin reconstruction method,
since we are concerned only with the top quark (or antitop quark) side.
What are relevant in our analysis are the effects of cuts and
acceptance corrections.
If we use the beamline basis, $E_{T}$ (transverse energy) 
and $|\eta|$ (pseudorapidity) cuts for leptons and jets 
strongly distort the double angular distribution
$d\Gamma/d\cos\theta_Wd\cos\theta_l$.
This is because small $E_{T}$ and large $|\eta|$ regions correspond
to the regions $\cos \theta_W \simeq \pm 1$ in this basis,
and events that fall into these kinematical regions are rejected.
In particular, a large part of the enhancement of the $f_2/f_1$ effects 
in the region $\cos \theta_W \sim -1$, $\cos \theta_l \sim -1$
is lost.
At Tevatron, the status of the off-diagonal basis is somewhat similar 
to the beamline basis, since the off-diagonal basis is not very different
from the beamline basis at Tevatron energies.
(At LHC, there is no good definition of the off-diagonal basis.)
The helicity basis turns out to be an optimal choice for our
purpose, since  this
basis points to every direction in the detectors.
After integrating over 
all top quark directions, effects of the cuts are averaged over and
no significant distortion from the original distribution is found.

Taking these into account, we define an effective spin direction by a 
projection of the lepton direction onto the helicity
basis:
\bea
\vec{S}_{\rm SH}  =
\mathrm{sign}(\cos\Theta) \times \frac{\vec{p}_t}{|\vec{p}_t|} 
\quad ,
\label{eq7}
\eea
where $\Theta$ is the angle between the charged lepton 
and the  top helicity direction $\vec{p}_t/|\vec{p}_t|$
(opposite of the antitop direction)
in the top rest frame.
We refer to the above effective spin direction as 
{\it signed-helicity} (SH) direction.

We conclude that the signed-helicity direction (\ref{eq7}) is a valid 
spin direction in studying the double angular distribution
$d\Gamma/d\cos\theta_Wd\cos\theta_l$.
It is also important that the dependence of the distribution on
the anomalous couplings is approximately reproduced in this spin
reconstruction method.

\section{Monte Carlo simulation}

In order to estimate the sensitivity for the anomalous couplings in the top 
decay, we perform Monte Carlo (MC) event generation and detector 
simulations. The events are produced with both Tevatron Run II ($p\bar{p}$ 
collisions in $\sqrt{s}$ = 1.96 TeV) and LHC ($pp$ collisions in 
$\sqrt{s}$ = 14 TeV) conditions.

The event generation for the $t\bar{t}$ signal samples
is implemented by GR@PPA event generator \cite{grappa} 
interfaced by PYTHIA v6.226 showering MC \cite{pythia}. 
The GR@PPA produces the hard process based on the $t\bar{t}$ matrix 
element calculation at the tree level. 
The whole decay chain of the top quark 
is included in the diagram calculation, so that the spin correlations in the 
top decay are fully reproduced. The anomalous couplings in the top decay are 
also included. 
PYTHIA performs fragmentation, parton showering, and hadronization.
On the other hand, underlying events are produced by PYTHIA alone,
using the parameters tuned to reproduce the Tevatron real data.

The detector simulation is performed by smearing energies for the stable 
particles deposited into a proper segmentation of the calorimeter geometry. 
The detector is assumed to stretch in absolute pseudorapidity ($|\eta|$) up to 
3.0 and be segmented by 0.1 $\eta$ bins and 15$^\circ \phi$ (azimuthal) bins.
The transverse energies $E_{T}$ for undecayed particles are summed up in each 
bin and are smeared by Gaussian distribution with 75\% $\times$ $\sqrt{E_{T}}$ 
standard deviation in GeV.\footnote{
We used the PYTHIA 
machinery to implement these resolution effects. 
}
As for leptons, we replace their measured momenta by the
values at the generator level.

Although our MC simulations are not fully realistic, we consider them to be  useful 
for giving rough estimates of 
the sensitivity to the anomalous couplings 
before performing full simulations. 
In particular, as for Tevatron experiments, our MC simulation would give
quite reasonable results.
On the other hand, as for LHC studies, there are some other 
important ingredients
that should be taken into account before giving more realistic
estimates of the sensitivity.
Among them, the most important effects would be those of extra jets events,
i.e.\ $t\bar{t} + n$-jets events, which are not included in our event generation.
Their effects are expected to be small at Tevatron.

A jet is clustered by {\tt PYCELL} routine in PYTHIA with cone size 0.4. 
We do not simulate $b$ tagging.
Instead a $b$-jet is identified as the nearest jet with the minimum separation 
$\Delta R$ between a jet and a $b$-quark at the generator level. The 
separation ($\Delta R$) is defined as 
\begin{equation}
\Delta R \; = \; \sqrt{\Delta \phi^{2} + \Delta \eta^{2}}
\quad ,
\label{eq8}
\end{equation}
where $\Delta \phi$ and $\Delta \eta$ are the separation in the azimuthal 
angle and the pseudorapidity for every pair of a jet and $b$-quark at the 
generator level, respectively. 

We select the $lepton+jets$ channel in the $t\bar{t}$ production process
by requiring to pass the cuts \\ \\
\hspace{10mm}
\begin{tabular}{lllll}
& Tevatron & & LHC \\
lepton~~~~~~ & $p_{T} \; \geq \; 20 \; \mathrm{GeV}$,
&  $|\eta| \; \leq \; 1.0$  ~~~~~
& $p_{T} \; \geq \; 20 \; \mathrm{GeV}$,
&  $|\eta| \; \leq \; 2.5$ 
\\
$b$--jet & $E_{T} \; \geq \; 15 \; \mathrm{GeV}$,
&  $|\eta| \; \leq \; 1.0$  ~~~~~
& $E_{T} \; \geq \; 30 \; \mathrm{GeV}$,
&  $|\eta| \; \leq \; 2.5$ 
\\
other jet & $E_{T} \; \geq \; 15 \; \mathrm{GeV}$,
&  $|\eta| \; \leq \; 2.0$  ~~~~~
& $E_{T} \; \geq \; 30 \; \mathrm{GeV}$,
&  $|\eta| \; \leq \; 2.5$ 
\\
& $E\sla_{T} \, \geq \; 20 \; \mathrm{GeV}$ & & 
$E\sla_{T} \, \geq \; 20 \; \mathrm{GeV}$ &
\end{tabular}
\label{eq9}
\\ \\
where $E\sla_{T}$ is the missing transverse energy calculated by the vector 
summation of the candidate lepton and four jets. We require two $b$-jets 
within at least 4 jets in each event.

The detection efficiency is 1.2\% including acceptance corrections
and branching fraction of $t\bar{t}$ events to the $lepton+jets$ channel,
with 25\% double $b$-tagging efficiency
at Tevatron condition;
the corresponding
detection efficiency is 2.3 \% with 60\% double $b$-tagging
efficiency at LHC condition.
Kinematical acceptance
fluctuates only within 0.5\% for various values of the anomalous couplings.

\section{Event reconstruction}

We measure the double angular distribution of $W$ and 
the charged lepton. For this purpose, it is important to reconstruct the full 
event topology of top quark events. 
The reconstruction of the $t\bar{t}$ 
event topology is performed by a likelihood method on event by event basis,
using the $lepton + jets$ events selected as above. 
Our kinematical likelihood event reconstruction is based on 
that of \cite{ikematsu}, 
which is a dedicated study for top quark reconstruction at future $e^+e^-$  linear 
colliders. 
In order to apply it to hadron collider experiments, some 
modifications are implemented. We assume that the energy-momentum of leptons 
and directions of jets can be measured accurately. Thus, for the 
$lepton + jets$ channel, we treat only 5 parameters out of 16 kinematical 
parameters as unknown. 
These 5 parameters are assigned to be the $E_{T}$ of 
jets and the boost vector of $t\bar{t}$ c.m.\ system along the beam axis.
We neglect the transverse momentum of the $t\bar{t}$ system.
The 
fit is constrained by the top quark mass, $W$ mass and parton 
distribution function (PDF). Our likelihood function is formed as 
\begin{equation}
L \; = \Bigl[ \prod_{i=1}^{4} P_{jet}^{i}(E_{T}^{exp.},E_{T}^{gen.}) \Bigr] \cdot
P_{\Gamma_{W^{+}}} \cdot P_{\Gamma_{W^{-}}} \cdot 
P_{\Gamma_{t}} \cdot P_{\Gamma_{\bar{t}}} \cdot P_{Z_{PDF}}^{t\bar{t}} \quad ,
\label{eq10}
\end{equation}
where $P_{jet}^{i}$ is the jet response function
that relates the measured jet energy
to the corresponding parton-level energy;
$P_{\Gamma_{W^{+}}}$, $P_{\Gamma_{W^{-}}}$, $P_{\Gamma_{t}}$ and 
$P_{\Gamma_{\bar{t}}}$ are Breit-Wigner functions\footnote{We fix the top 
quark pole mass to be 178~GeV.} for $W^{+}$, $W^{-}$, $t$ and $\bar{t}$, 
respectively;
$P_{Z_{PDF}}^{t\bar{t}}$ 
constrains the boost momentum of the $t\bar{t}$ c.m.\
system by PDF function. 
Whenever more than one solution for jet assignment is found 
in each $t\bar{t}$ candidate event,
we take the one that maximizes the likelihood function.

The jet response functions are constructed for the 
light-quark jets from $W$ and for the $b$-jets separately
in the following manner.
A single 
parton generator is used to estimate them. 
A parton with a 
certain fixed energy and direction is generated and
passed through the detector simulation. 
The energy of the
reconstructed jet at the detector level is fitted by Gaussian distribution and
is defined to be the probability density for the given energy 
and direction of the parent parton. 
We iterate this procedure with various energies and directions, 
and
define jet energy scales 
of the responses in the calorimeter positions as the jet response 
functions. 

\begin{figure}[tb]
\begin{center}
\begin{tabular}{ccc}
\includegraphics[width=6.0cm]{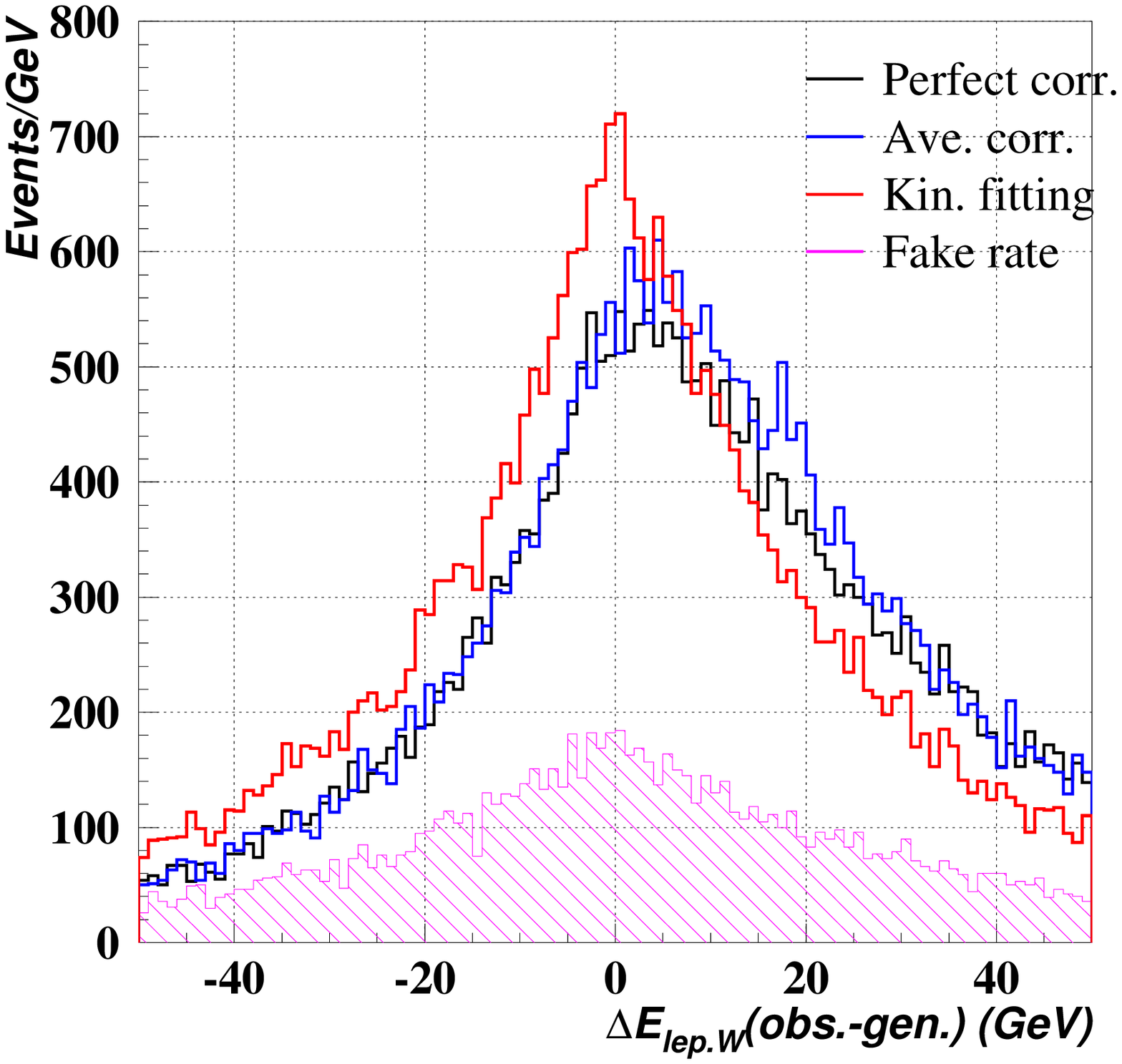} &&
\includegraphics[width=6.0cm]{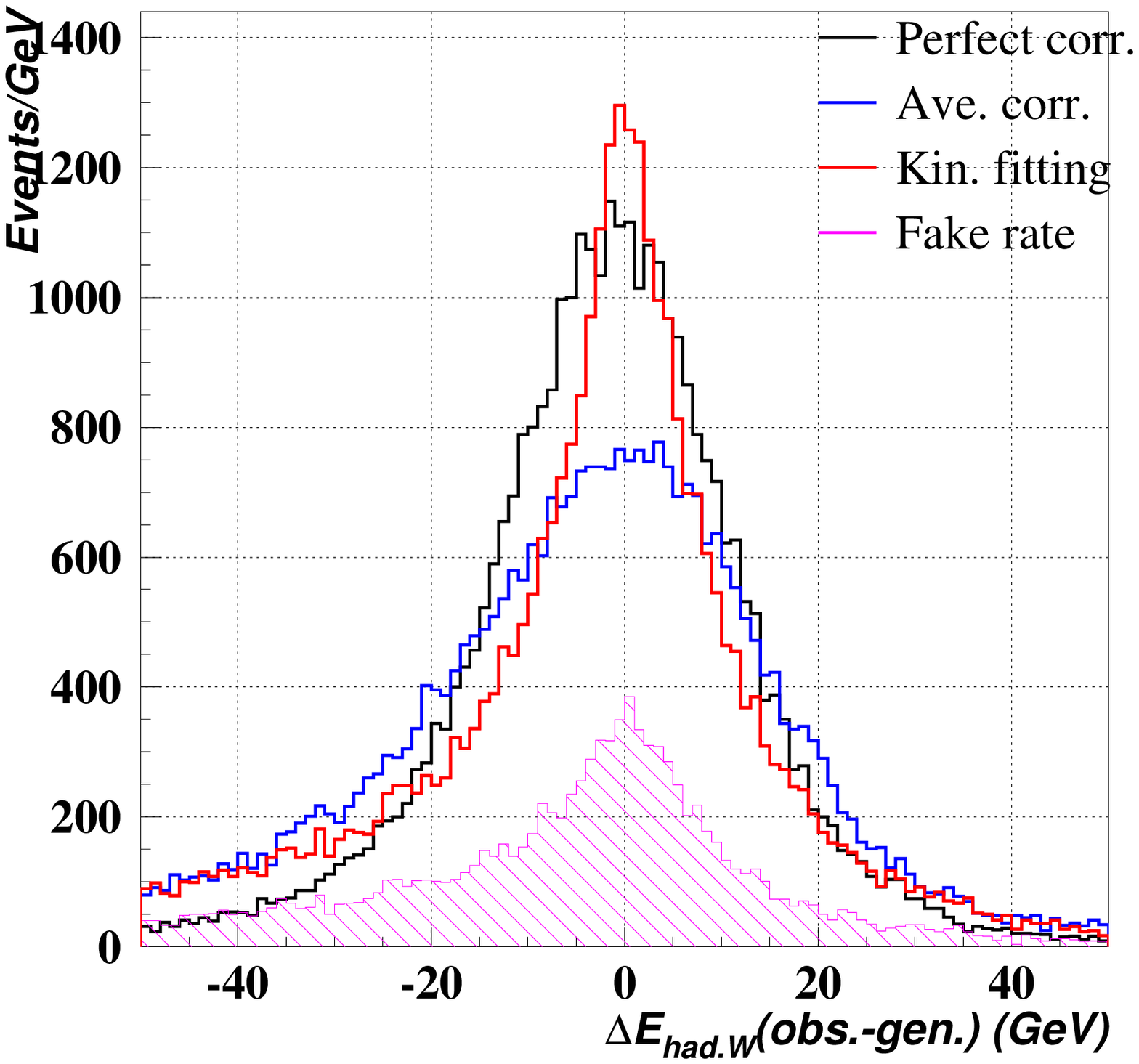}\\
(a) && (b)\\ \\
\includegraphics[width=6.0cm]{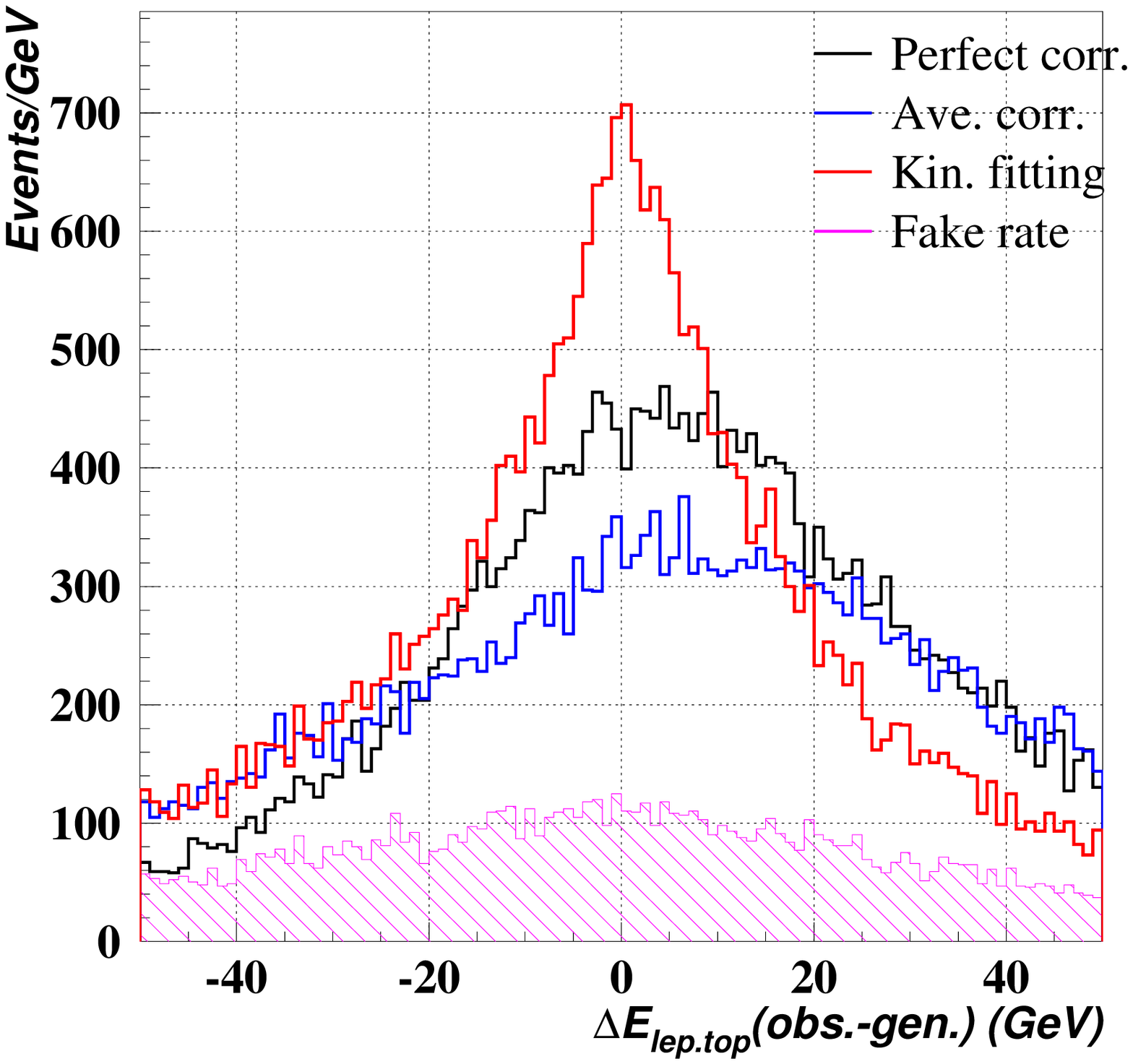} &&
\includegraphics[width=6.0cm]{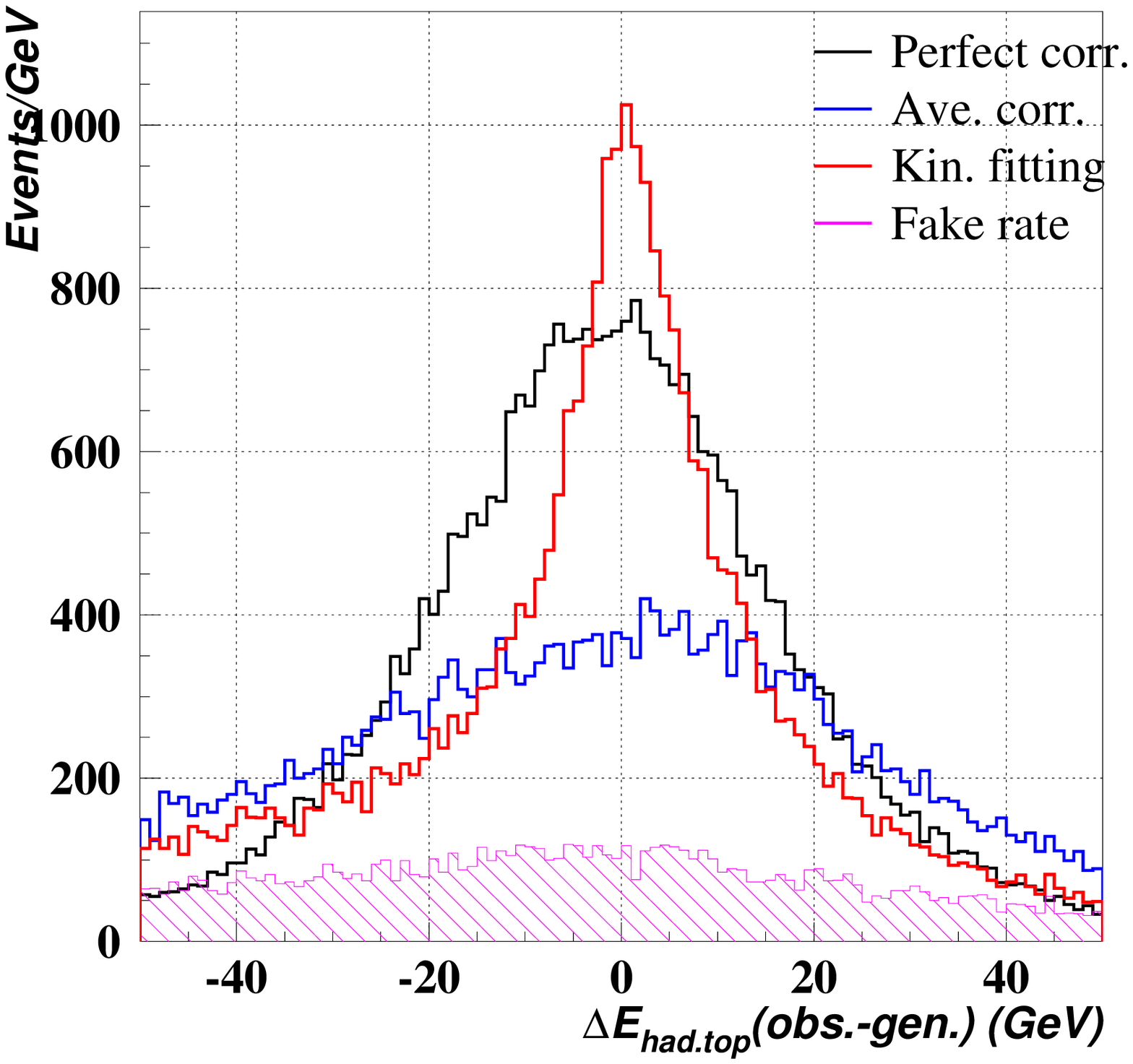}\\
(c) && (d)\\
\end{tabular}
\caption{\small Deviations between the 
parton energies at the generator level and corrected jet 
energies at the detector level for (a) leptonically decayed $W$ boson,
(b) hadronically decayed $W$ boson,
(c) leptonically decayed top quark, and
(d) hadronically decayed top quark.
}
\label{de_partons}
\end{center}
\end{figure}

Let us demonstrate how well 
the parton kinematics are reconstructed.
In Figs.~\ref{de_partons}
we show the difference between the parton energy
at the generator level and the energy
determined by the 
fit at the detector level:
we show the energy differences
for (a) the leptonically-decayed $W$ 
boson,
(b) the hadronically-decayed $W$ boson, 
(c) the leptonically-decayed top quark, and 
(d) the hadronically-decayed top quark, respectively.
The fake contributions are also shown as hatched regions, 
which correspond to the events that include jets not 
assigned in correct combinations by the fit.
(We define that jets are correctly assigned, if the directions of 
all jets are matched with those of the original partons within jet cone radius of 
0.4.)
For comparison, the other 
reconstruction methods ``perfect correction'' and ``average correction'' are 
also presented in the figures. 
With the ``perfect correction'', the jet 
energy is defined to be the corresponding parton energy at the generator
level
smeared with a finite 
detector resolution. 
With the ``average correction'', the jet energy is 
uniformly corrected by the jet energy
scale for the mean value.\footnote{
Jet energy reconstruction methods similar to the  ``average correction'' is 
being used in current studies on
LHC experiments.
} 
Note that the fake contributions are not included for these two correction 
methods. 

From Figs.~\ref{de_partons}, we can see that the likelihood fitting method 
reproduces the event kinematics considerably better than the other 
two methods. 
Quality of the energy
reconstructions of $W$ and top quark is worse on the leptonic side
than on the hadronic side. 
This follows from a poorer
resolution for the neutrino momentum on the leptonic side, 
which is defined as the opposite of the vector 
summation of all (four) jets and lepton in the $t\bar{t}$ c.m.\ frame.
On the other hand, $W$ and top 
quark on the hadronic side are reconstructed using the two and three jets,
respectively. 
One sees that, 
in the tail regions of the distributions in the figures, 
the correctly assigned events are suppressed,
which also shows that the likelihood fitting works as expected.

Using the reconstructed momenta of $t$ and $\bar{t}$, we
reconstruct the effective top spin according to the signed-helicity method
Eq.~(\ref{eq7}) as follows.
The top helicity axis is defined in the top quark 
rest frame as (the opposite of) the direction
of the 
momentum of the hadronically-decayed antitop quark, which sequentially decayed into 
three jets. 
The sign of the top spin is defined by the direction of the 
charged lepton in the top rest frame.
The reconstructed top quark momentum is also 
used to measure the helicity angle of the charged lepton, since the original 
direction of $W$ in the $W$ rest frame is equivalent 
to the opposite of the leptonically-decayed top quark direction
in the $W$ rest frame
(see Sec.~\ref{s2}).

\section{Sensitivity study}

In this section, we study the sensitivity for the anomalous couplings using 
the $lepton + jets$ events reconstructed by the kinematical likelihood fitting. 

\begin{figure}[t]
\begin{center}
\begin{tabular}{ccc}
\includegraphics[width=6.0cm]{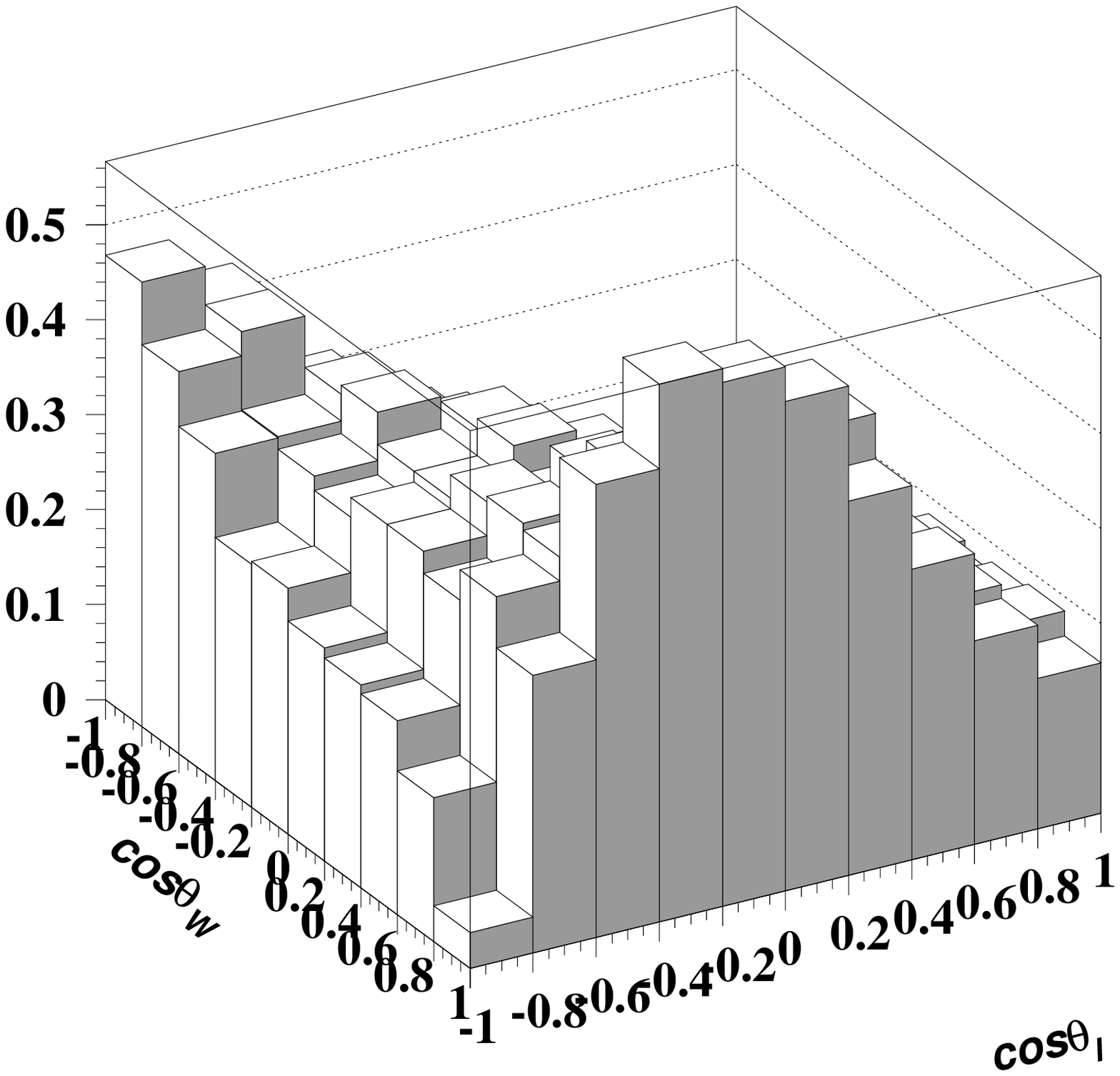} &&
\includegraphics[width=6.0cm]{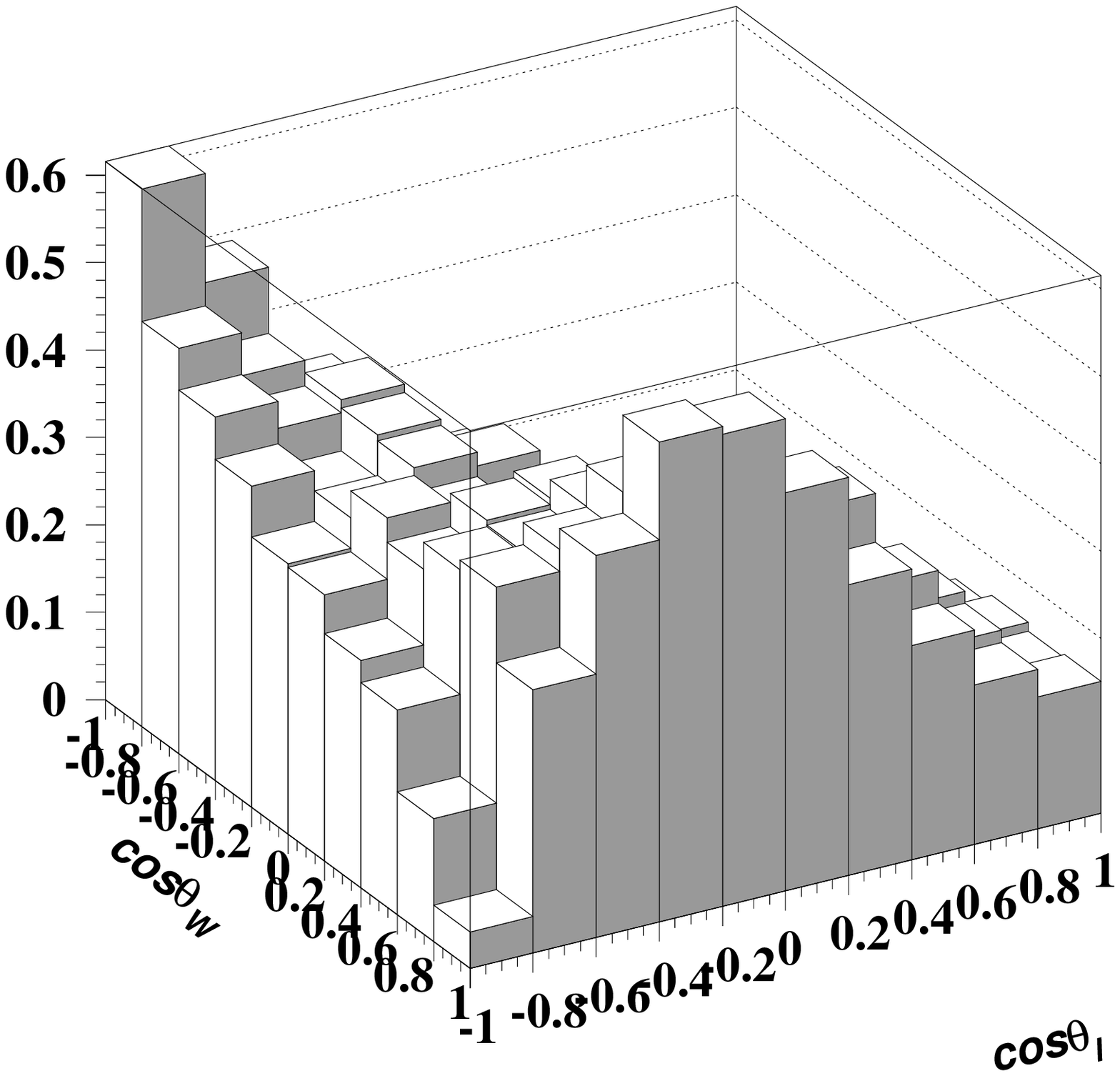}\\
(a) && (b)
\end{tabular}
\caption{\small Normalized differential decay distributions 
using the signed-helicity direction {\it after event reconstruction
and kinematical cuts} 
(a) for $(f_1,f_2) =(1,0)$,
and (b) for $(f_1,f_2) =(1,0.3)$.
}
\label{det_spin}
\end{center}
\end{figure}

We show in Figs.~\ref{det_spin} the double angular distributions
$d\Gamma/d\cos\theta_W\cos\theta_l$ using the
MC events,
after event selection and event reconstruction by the kinematical likelihood 
fitting.
Compare with the corresponding parton distributions 
at the generator level in Figs.~\ref{fin_smgen}.
One can see that, even after cuts, the dependence on the anomalous
couplings remains in
the $W_{T}$ region ($\cos\theta_{W}\sim-1$, $\cos\theta_{l}\sim-1$).

\begin{figure}[t]
\begin{center}
\includegraphics[width=6.0cm]{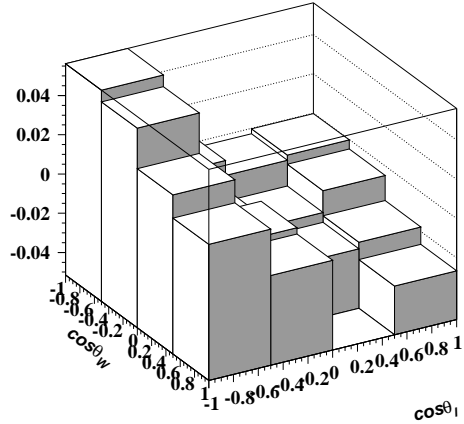}
\caption{\protect \small 
Deviation of the normalized angular distribution when the anomalous couplings are
varied from $(f_1,f_2) =(1,0)$ to
 (1, 0.3).}
\label{dif_spin_sm_f103}
\end{center}
\end{figure}
The difference between the angular distributions corresponding
to the anomalous couplings 
($f_{1}$,$f_{2}$) = (1,0.3) and (1,0) is shown in 
Fig.~\ref{dif_spin_sm_f103}. 
The difference is maximized in the $W_{T}$ region 
($\cos\theta_{W}\sim-1$, $\cos\theta_{l}\sim-1$) and minimized in its diagonal 
opposite
region ($\cos\theta_{W}\sim1$, $\cos\theta_{l}\sim1$). 
The other two (diagonal)
regions have weaker dependences on the anomalous couplings. 
When signal 
statistics is small or the background contribution is not 
well-understood, a simple but not elaborate
method to determine the anomalous couplings would be
practical for a first analysis. 
Hence, we divide the kinematical region into 4 regions and
simply count the number of events in each region.
The regions are defined as follows:
\begin{equation}
\begin{array}{lrccrcl}
\mathrm{Region\ A} : \quad & -1 \; & \leq \; \cos\theta_{W} \; \leq \; 0 
& \quad \mathrm{and} \quad & -1 \; & \leq \; \cos\theta_{l} \; \leq \; 0 & \\
\mathrm{Region\ B} : \quad & -1 \; & \leq \; \cos\theta_{W} \; \leq \; 0
& \quad \mathrm{and} \quad & 0 \; & \leq \; \cos\theta_{l} \; \leq \; 1 & \\
\mathrm{Region\ C} : \quad & 0 \; & \leq \; \cos\theta_{W} \; \leq \; 1 
& \quad \mathrm{and} \quad & -1 \; & \leq \; \cos\theta_{l} \; \leq \; 0 & \\
\mathrm{Region\ D} : \quad & 0 \; & \leq \; \cos\theta_{W} \; \leq \; 1 
& \quad \mathrm{and} \quad & 0 \; & \leq \; \cos\theta_{l} \; \leq \; 1 & \quad .\\
\end{array}
\label{eq11}
\end{equation}

\begin{figure}[t]
\begin{center}
\includegraphics[width=8.0cm]{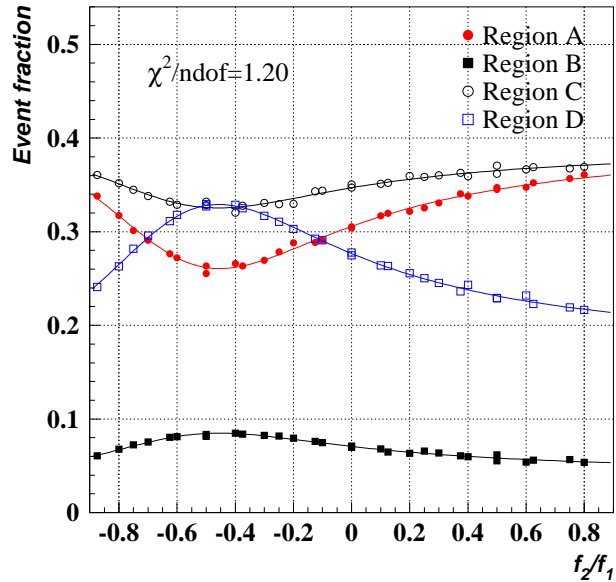}
\caption{\small 
MC data and fitting results of the event fractions in 
each region  as functions of $f_{2}$/$f_{1}$.
Each region is defined in Eq.~(\ref{eq11}). 
}
\label{ratiofit}
\end{center}
\end{figure}

The dependences of
the event fractions in these regions on the anomalous couplings
are shown in Fig.~\ref{ratiofit}.
The regions A and D are the regions most 
sensitive to the anomalous couplings, while the regions B and C 
are less sensitive regions. 
We can see that the event fraction in region A 
increases with $f_{2}$/$f_{1}$ when $f_{2}$/$f_{1}$ $>$ 0, and 
takes a minimum value around $f_{2}$/$f_{1} \approx -0.45$, and then 
increases again if we lower $f_{2}$/$f_{1}$ below $-0.45$. 
The event fraction in region D 
has an opposite behavior to that of region A. 
All the event fractions take maximum or minimum values around
$f_{2}$/$f_{1} = -M_W/M_t \approx -0.45$, where the transverse 
component of $W$ is canceled; cf.~Eq.~(\ref{doubleangdistr}).

We fit the MC data (shown by discrete points in Fig.~\ref{ratiofit})
by analytic functions as follows.
In an ideal case, we can integrate Eq.~(\ref{eq15}) analytically
over each of the regions A--D.
The event fraction distributed in 
each region takes a form
\begin{equation}
F^i(x) \; = \; \frac{a^i_{1}x^{2} \; + \; 
a^i_{2}x \; + \; a^i_{3}}
{\widetilde{\Gamma}_{\rm tot}(x)} \quad ,
~~~~~
x=\frac{f_{2}}{f_{1}} ,
~~~~~
i={\rm A,B,C,D}
\label{eq12}
\end{equation}
where 
${\widetilde{\Gamma}_{\rm tot}(x)} =\Gamma_{t \to bW}(f_1,f_2)/|f_{1}|^{2} $.
($\Gamma_{t \to bW}(f_1,f_2)$ is the partial decay width of the top quark.)
Parameters $a^i_1, a^i_2, a^i_3$ can be expressed analytically in terms of
the top and $W$ masses.
Note that 
since $f_{1}$ only contributes to the normalization of the differential 
angular distribution which does not affect the shape of the distribution, the 
event fractions depend only on $x=f_{2}/f_{1}$ regardless of 
various choices of $f_{1}$ and $f_{2}$. 

In a realistic case, the distributions are affected by the finite
resolution of detectors, by cuts, by fake contributions, etc.
Here,
we fit the event fractions generated by MC simulation with high statistics
by the same functional form as in Eq.~(\ref{eq11}), taking
$a^i_1, a^i_2, a^i_3$ as the parameters to be determined by the fit.
The sum of the event fractions in four regions is normalized to one,
so that 9 parameters are decided by 
minimizing the fitting $\chi^{2}$. 
The MC data and fitting results of the event fractions in 
each region are shown as functions of $f_{2}$/$f_{1}$ in 
Fig.~\ref{ratiofit}. 
The $\chi^{2}$ minimum per each degree of freedom
takes a reasonable value $\approx 1.20$.
The functions $F^i(x)$ determined by the fit are used to estimate sensitivity 
to the anomalous couplings. 

In Table \ref{tablimit}, the expected 
bounds on the coupling ratio
at 95\% C.L.\ are shown, corresponding to 100 and 1000 selected
events (after cuts)
for the Tevatron experiment and 100k selected events (after cuts)
for the LHC experiment,
respectively.\footnote{
Using the detection efficiencies estimated at the end of Sec.~4, 
100 and 1000 double $b$-tagged
events at Tevatron are translated roughly
to 1 and 10~fb$^{-1}$ integrated luminosities,
respectively, and 100k events to 10 fb$^{-1}$ at LHC.
} 
Input parameters of the 
MC simulations are taken as $(f_1,f_2)=(1,0)$ (tree-level SM values). 
Only 
statistical errors are taken into account to obtain the allowed regions. 
For 
comparison,  we present the allowed regions using an ideal
off-diagonal direction (for Tevatron), in which the spin
direction is reconstructed using the
off-diagonal basis with the sign ambiguity
resolved by looking into the information at the generator level;
we may consider that this ideal off-diagonal 
direction approximates the true spin direction well, so that
the corresponding results can be used as references
(although these include 
effects of kinematical cuts  as well as contamination by fake events). 
We also present the allowed regions using only the events with 
correct assignment of two $b$-jets using the signed helicity direction.

In Table \ref{tablimit}, the bounds using 
the signed helicity direction are not very different from those
using the ideal off-diagonal spin 
direction at Tevatron. 
Since the latter results can be regarded as references for optimal reconstruction
of the top spin, it is seen that the signed helicity direction is quite efficient
for this analysis.\footnote{
In \cite{letter}, the sensitivity using the effective spin direction is estimated
to be about half of that using the true spin direction, if we ignore 
experimental environment.
Here, we have shown that at Tevatron 
the effects of kinematical cuts 
are quite large if we use the idal off-diagonal spin 
basis (true spin direction), so that
after the cuts,
the sensitivities are not very different whether we use
the signed-helicity direction or
the idal off-diagonal spin basis.
}
In addition, the sensitivities can be 
improved if we can remove misassignment of
the $b$-jets.

Although it is obvious that the expected bound becomes tighter as the number of 
events increases, the way the bound shrinks with statistics is rather peculiar.
This is because the event fractions have
characteristic (non-linear)
$f_2/f_1$-dependences, almost symmetric under reflection
with respect to $f_2/f_1 \approx -0.45$; see Fig.~\ref{ratiofit}.
At low statistics (100 events or less), the bound on $f_2/f_1$ is fairly loose.
When the statistics is increased, the bound does not simply scale with $1/\sqrt{N}$ but
becomes narrower much faster, with 
a two-fold ambiguity that remains,
i.e., the regions around $f_2/f_1 \approx 0$ and $f_2/f_1 \approx -0.75$
cannot be discriminated.
Once the number of events 
exceeds a few hundred, the bound scales with $1/\sqrt{N}$,
since the dependences of the event fractions can be approximated by linear
responses.
This gives a motivation to increase the number of events (after cuts) 
at least beyond a few hundred at Tevatron experiment.

\begin{figure}[t]
\begin{center}
\includegraphics[width=8.0cm]{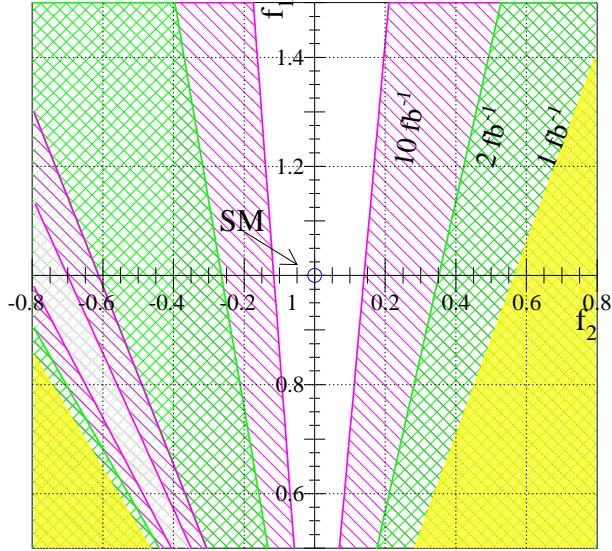}
\caption{\small 
Expected
excluded regions at 95\% C.L.
in the $(f_2,f_1)$-plane at Tevatron. 
The shaded regions correspond to
1 fb$^{-1}$, 2 fb$^{-1}$, and 10 fb$^{-1}$ 
integrated luminosities, respectively. 
The input SM point is located at ($f_{2}$,$f_{1}$)=(0,1).}
\label{limit_dsh}
\end{center}
\end{figure}

Finally, we show the expected excluded regions 
in the $(f_2,f_1)$-plane at 95\% C.L.
for the Tevatron case in Fig.~\ref{limit_dsh}. 
We  anticipate that our 
method allows us to cover a wide region in the parameter space even in this 
simplified counting experiment.

\begin{table}
\begin{center}
\begin{tabular}{c|c|c|c}
& \multicolumn{2}{c|}{Tevatron ($\sqrt{s}=1.96$ TeV)}
& \multicolumn{1}{c}{LHC ($\sqrt{s}=14$ TeV)} \\ \cline{2-4}
Number of events
& 100 & 1000 & 100k \\
\hline 
\hline 
Signed-helicity direction
& $-0.93$ $<$ $\frac{f_{2}}{f_{1}}$ $<$ 0.57
& $-0.12$ $<$ $\frac{f_{2}}{f_{1}}$ $<$ 0.14,
& $-0.01$ $<$ $\frac{f_{2}}{f_{1}}$ $<$ 0.01, \\
& 
& $-0.81$ $<\frac{f_{2}}{f_{1}}<$ $-0.70$
& $-0.74$ $<\frac{f_{2}}{f_{1}}<$ $-0.72$ \\ \hline
Ideal off-diagonal direction
& $-0.84$ $<$ $\frac{f_{2}}{f_{1}}$ $<$ 0.50
& $-0.11$ $<$ $\frac{f_{2}}{f_{1}}$ $<$ 0.12,
& Not applicable \\
& 
& $-0.73$ $<\frac{f_{2}}{f_{1}}<$ $-0.61$
&  \\ \hline
Signed-helicity direction
& $-0.29$ $<$ $\frac{f_{2}}{f_{1}}$ $<$ 0.39,
& $-0.09$ $<$ $\frac{f_{2}}{f_{1}}$ $<$ 0.10,
& $-0.01$ $<$ $\frac{f_{2}}{f_{1}}$ $<$ 0.01, \\
with correct $b$ assignment
& $-0.89$ $<\frac{f_{2}}{f_{1}}<$ $-0.59 $
& $-0.80$ $<\frac{f_{2}}{f_{1}}<$ $-0.71 $
& $-0.75$ $<\frac{f_{2}}{f_{1}}<$ $-0.74 $ \\ \hline 
\end{tabular}
\caption{\small 
Expected bounds at 95\% C.L.\ corresponding to 100 and 1000 
events (after cuts)
for Tevatron and 100k events (after cuts) for LHC. 
Input parameters of the 
MC simulations are taken as $(f_1,f_2)=(1,0)$. 
Only statistical errors are taken into account. 
For 
comparison, the bounds using an ideal off-diagonal direction, and those
using only the events with 
correct assignment of two $b$-jets in the signed helicity method are presented.
}
\label{tablimit}
\end{center}
\end{table}

\section{Conclusions}

We have studied sensitivities to the top quark decay anomalous couplings
$f_1$ and $f_2$ at hadron colliders,
taking into account realistic experimental conditions
expected at Tevatron and LHC.
Since large samples of top quarks are expected at these colliders,
and since the decay processes of the top quark can be predicted reliably
by perturbative QCD, we can achieve high sensitivities to the anomalous
couplings from detailed studies of the top quark decay processes.
We used a likelihood method to fully reconstruct the momenta of
the partons in the {\it lepton+4 jets} mode.
Furthermore, we devised a new method for reconstructing an effective spin
direction Eq.~(\ref{eq7})
(referred as {\it signed-helicity direction}) of the leptonically-decayed top quark.
It is defined as the projection of the lepton direction onto the
top helicity axis in the top rest frame.
This method does not require 
reconstruction of the spin
of the hadronically-decayed top quark, hence it helps to elude
possibly large systematic uncertainties.
These two techniques, when used in combination, revealed to be quite powerful
for the sensitivity study.

We analyzed a double angular distribution
$d\Gamma(t\to b l \nu)/d\cos\theta_W d\cos\theta_l$.
The $W_T$ region $\cos\theta_W \sim -1$, $\cos\theta_l \sim -1$ 
of the distribution is 
sensitive to the ratio of the anomalous couplings
$f_2/f_1$.
We confirmed that this feature is preserved even after kinematical cuts.
We note that if we choose a spin axis other than top helicity basis, 
such as beamline basis or 
off-diagonal basis, sensitivity to $f_2/f_1$ is substantially reduced due to 
effects by the kinematical cuts.

In order to give reliable estimates, we developed an event 
generator incorporating in the matrix element proper spin correlations of partons 
as well as the anomalous
couplings in the top decay vertices.
We also simulate the detector effects by 
assuming a simple geometry and energy resolutions based on the CDF and ATLAS 
detectors for Tevatron and LHC experiments, respectively. 
After event selection,
the event kinematics are reconstructed by the kinematical 
likelihood fitting on an event by event basis. 
It not only improves the jet energy scale from the measured jet energy to the 
corresponding parton energy but also helps to select the correct configuration of 
the jets in the top event topology. 
Furthermore, the likelihood fitting method 
improves reconstruction of
the hadronically-decayed top quark's energy and momentum, 
which is directly reflected to
the determination of the top spin 
direction as well as lepton helicity angle.

As a first analysis, we simply counted the event fractions of the double
angular distribution divided into 4 regions. 
Then we performed $\chi^2$--fits to these event fractions
in order to find sensitivities
to $f_2/f_1$.
The results can be summarized as follows.
The bounds obtained at 95\% C.L. read
\bea
\begin{array}{cl}
-0.93 < \frac{f_2}{f_1} < 0.57 & \mbox{for 100 reconstructed events at Tevatron,} \\
-0.81<\frac{f_2}{f_1}<-0.70, ~~ -0.12<\frac{f_2}{f_1}<0.14
& \mbox{for 1000 reconstructed events at Tevatron,} \\
-0.74 < \frac{f_2}{f_1}<-0.72, ~~-0.01 < \frac{f_2}{f_1} < 0.01
& \mbox{for 100k reconstructed events at LHC}. 
\end{array}
\label{eq13}
\eea
We took into account only the statistical errors and neglected
systematic errors.
Due to characteristic dependences of the event fractions on $f_2/f_1$, the
bound on $f_2/f_1$ shrinks quickly as the number of top quark
events increases up to a few hundred.
For more events, the bound scales with $1/\sqrt{N}$, and there remains a two-fold
ambiguity for the allowed ranges of $f_2/f_1$.

Although some simplifications have been made, we consider that our MC study 
for the Tevatron experiment imitates
realistic experimental conditions closely enough to give reasonable estimates
for the sensitivities to the anomalous couplings.
On the other hand, as for the LHC case, some important ingredients 
are still missing in the MC simulation
(the most important one would be $t\bar{t}+n$-jets events), 
so our results should be taken as first rough estimates.

Since our methods for event reconstruction and effective top spin reconstruction 
are fairly simple, we would expect that they can be applied to other
analyses, such as precise determination of $W$ polarization states in top
decay.

\section*{Acknowledgments}

One of the authors (S.T.) is grateful to Minami-Tateya numerical calculation group for 
collaborations in developing 
the event generator incorporating the top anomalous couplings.

\newpage

\section*{Appendix: Decay angular distribution using the signed-helicity 
spin direction}

In this Appendix, we present the analytic formula for the double angular
distribution in the decay of top quark,
$d\Gamma/d\cos\theta_W d\cos\theta_l$, when we use the
lepton direction projected onto any spin basis
for the reconstructed top quark
spin direction.
(When we choose the helicity basis, we call the
reconstructed spin direction as ``signed-helicity'' direction.)
We assume the parent top quark to be unpolarized and neglect the
effects of kinematical cuts.

An arbitrary unit vector $\vec{n}$ is chosen as the spin axis
in the top rest frame.
Then, if ${\vec{n}\cdot\vec{p}_l}>0$, we define the
``spin vector'' to be $\vec{n}$ whereas, if ${\vec{n}\cdot\vec{p}_l}<0$,
we define the
``spin vector'' to be $-\vec{n}$.
The differential decay distribution 
$d\Gamma/d\cos\theta_Wd\cos\theta_l$
with respect to thus defined ``spin vector''
can be computed analytically
as follows.
The angle $\Theta$ between $\vec{n}$ and $\vec{p}_l$
is given by
\bea
&&
\cos \Theta \equiv
\frac{\vec{n}\cdot\vec{p}_l}{|\vec{p}_l|}=
\frac{\sqrt{1-\beta_W^2}}{1+\beta_W \cos\theta_l}
\left(
\sin \theta_l \cos \phi_l \sin \theta_W
+ \frac{\cos \theta_l + \beta_W}{\sqrt{1-\beta_W^2}} \cos\theta_W
\right) ,
\\ &&
\beta_W=\frac{M_t^2-M_W^2}{M_t^2+M_W^2} ,
\eea
where $\theta_W$ and $\phi_l$ are defined as in Fig.~\ref{angle_def}
with respect to $\vec{n}$.
Since the definition of the spin vector is reversed when $\cos \Theta <0$, in this case we need
to redefine the angles to be 
$\theta_W \to \pi - \theta_W$ and $\phi_l \to \phi_l + \pi$
in the cross section formula.
Therefore, noting that the initial top quark is unpolarized, 
the double angular distribution is given by
\bea
&&
\left[
\frac{d\Gamma(t \rightarrow bW \! \rightarrow  bl\nu)}
{d\cos\theta_{W}d\cos\theta_{l}} 
\right]_{\rm SH} 
\nonumber\\
&&
= 
\int_{\cos\Theta >0} \!\!\!\!\!\! d\phi_l \, 
\left[
\frac{d\Gamma(t \rightarrow bW \! \rightarrow  bl\nu)}
{d\cos\theta_{W}d\cos\theta_{l}d\phi_{l}} 
\right]_{\rm unpol.}
+
\left[
\int_{\cos\Theta <0} \!\!\!\!\!\! d\phi_l \, 
\left[
\frac{d\Gamma(t \rightarrow bW \! \rightarrow  bl\nu)}
{d\cos\theta_{W}d\cos\theta_{l}d\phi_{l}} 
\right]_{\rm unpol.}
\right]_{\begin{array}{l}\scriptstyle \theta_W \to \pi - \theta_W\\
\scriptstyle  \phi_l \to \phi_l + \pi
\end{array}}
\nonumber\\
&&
=
\left[
\frac{d\Gamma(t \rightarrow bW \! \rightarrow  bl\nu)}
{d\cos\theta_{W}d\cos\theta_{l}d\phi_{l}} 
\right]_{\rm unpol.} \times 2\, g(y)
,
\label{eq15}
\eea
where
\bea
&&
y = -\frac{\cos \theta_l + \beta_W}{\sqrt{1-\beta_W^2}}\,
\frac{\cot \theta_W}{\sin \theta_l} ,
\\
&&
g(x) = \left\{
\begin{array}{ll}
0&\mbox{if $x \geq 1$}\\
2\pi & \mbox{if $x \leq -1$}\\
\pi - 2 \arcsin x & \mbox{if $-1<x<1$}
\end{array}
\right. .
\eea
The decay distribution from an {\it unpolarized} top quark is given by
\bea
\left[
\frac{d\Gamma(t \rightarrow bW \! \rightarrow  bl\nu)}
{d\cos\theta_{W}d\cos\theta_{l}d\phi_{l}} 
\right]_{\rm unpol.} = 
\frac{1}{4}\,A \,
\biggl[
\Bigl(f_1\frac{M_t}{M_W}+f_2 \Bigr)^2
\, \sin^2 \theta_l
+ 4 \, \Bigl(f_1 +f_2\frac{M_t}{M_W} \Bigr)^2
\,\sin^4  \frac{\theta_l}{2} 
\biggr] \, .
\eea
It is independent of $\theta_W$ and $\phi_l$, since there is
no reference spin vector.

\end{document}